\documentclass[amsmath,amssymb,aps,10pt,pra,nofootinbib,twocolumn,showpacs]{revtex4-1}

\usepackage{graphicx}
\usepackage{grffile} 

\usepackage{bbold}
\usepackage{color}
\definecolor{gray}{rgb}{0.5,0.5,0.5}
\usepackage[utf8]{inputenc}
\usepackage{textcomp}
\usepackage{amsmath}

\newcommand{\ksr}{\underline{K}^\mathrm{sr}}
\newcommand{\ssr}{\underline{S}^\mathrm{sr}}
\newcommand{\ksrtilde}{\underline{\tilde{K}}}

\newcommand{\eq}[1]{Eq.~(\ref{#1})}
\newcommand{\fig}[1]{Fig.~\ref{#1}}
\newcommand{\onehalf}{\frac{1}{2}}

\graphicspath{{./figures/}}

\begin{document}
\title{Statistical Aspects of Ultracold Resonant Scattering}

\date{\today}

\author{Michael Mayle}
\author{Brandon P. Ruzic}
\author{John L. Bohn}
\affiliation{JILA, University of Colorado and National Institute of Standards and Technology, Boulder, Colorado 80309-0440, USA}

\date{\today}

\begin{abstract}
Compared to purely atomic collisions, ultracold collisions involving molecules have the potential to support a much larger number of Fano-Feshbach resonances due to the huge amount of ro-vibrational states available. In order to handle such ultracold atom-molecule collisions, we formulate a theory that  incorporates the ro-vibrational Fano-Feshbach resonances in a statistical manner while treating the physics of the long-range scattering, which is sensitive to such things as hyperfine states, collision energy and any applied  electromagnetic fields, exactly within multichannel quantum defect theory. Uniting these two techniques, we can assess the influence of highly resonant scattering in the threshold regime, and in particular its dependence on the hyperfine state selected for the collision. This allows us to explore the onset of Ericson fluctuations in the regime of overlapping resonances, which are well-known in nuclear physics but completely unexplored in the ultracold domain.
\end{abstract}

\maketitle

\section{Introduction}

Resonances have always played a key role in scattering experiments across many areas of physics, serving to nail down our understanding of the interaction between the collision partners. They play an additional role in dilute, ultracold atomic and molecular gases, where resonance positions can be moved relative to the (essentially zero) collision energy by means of applied electromagnetic fields. This circumstance allows one to control collision cross sections, as well as mean-field interactions in quantum degenerate gases. Dozens of magnetic-field Fano-Feshbach resonances have been identified and characterized in ultracold collisions of various alkali atoms \cite{RevModPhys.82.1225}; many are now working tools for research in many-body quantum physics. In the case of cold collisions of alkali atoms, the resonant states differ from the incident scattering states by the change of an internal spin. For this reason, the number of resonant states remains typically small and the resonances themselves usually remain well-separated and tractable.

This situation can be different, however, for collisions involving cold molecules, where rotational and vibrational excitations can also contribute to resonant states. Many such resonances have been predicted in theoretical treatments of cold molecular scattering \cite{PhysRevA.64.052703,PhysRevLett.89.203202,Krems2005,tscherbul:083201,Simoni2006,PhysRevA.77.030704,krems08,1367-2630-11-5-055021,simoni:032701,PhysRevA.79.062708,PhysRevA.84.042703}. While the number of resonances naturally grows in this case, nevertheless the individual resonances are typically well-resolved and manageable in number. This is particularly evident in cold collisions of molecules with helium atoms, relevant to buffer gas cooling, where light masses and shallow potential energy surfaces conspire to keep the density of resonant states low \cite{PhysRevA.62.032701,PhysRevLett.106.073201}. Resonances appear to be resolved even in collisions involving  light objects other than helium such as O$_2$ \cite{PhysRevA.64.052703}, Rb+OH \cite{PhysRevA.75.012704}, N+NH \cite{PhysRevLett.106.053201}, or Mg+NH \cite{PhysRevA.84.052706}. In  relatively ``clean'' systems like these, there remains hope of explicitly identifying the quantum numbers of resonances, and using them to back out accurate potential energy surfaces (PES). Indeed, energy resolution afforded at ultra low temperatures may allow for the elucidation of van der Waals \cite{Bala04,Weck2006} or transition state \cite{Truhlar1996,skodje2000} resonances, important for unraveling chemical reactions when a barrier is present.

There remains, however, a class of heavier molecules that have been or will be produced at ultracold temperatures.  Notable among these, and the subject of this paper, are diatomic species consisting of pairs of alkali atoms.  When such a molecule collides with another alkali atom, the PES is sufficiently deep that tens of vibrational levels, and hundreds of rotational levels, may be energetically accessible. In this case the density of resonant states (DOS) may be so high that individual resonances may not even be resolved, let alone identified. In such a case, it would be worthwhile to understand the effect of all these resonances on observed collision cross sections.

Theories relating to high-DOS scattering have long ago been formulated, notably in chemistry and in nuclear physics. On the one hand, the theory of unimolecular dissociation regards the problem in the time domain. If a polyatomic molecule is given enough energy to break a particular bond, say by absorbing an appropriate photon, it does not necessarily immediately dissociate. Rather, it can lose energy in many irrelevant degrees of freedom until, by accident, sufficient energy lands in the desired bond to break it. The theory of this process, known as the Rice-Ramsperger-Kassel-Marcus (RRKM) theory \cite{levinebook}, expresses the mean rate of dissociation as
\begin{eqnarray}
\label{RRKM_rate}
k_{\rm RRKM} = \frac{ 1 }{ 2 \pi } \frac{ N_a }{ \hbar \rho }.
\end{eqnarray}
Here $\rho$ represents the (very large) density of resonant states, while $N_a$ represents the (small) number of quantum states available at the transition state which lead to dissociation.

On the other hand, scattering experiments in nuclear theory have inspired statistical ideas of highly resonant scattering more in the energy domain. Again, a high density of states is expected because of the many strongly interacting nucleon degrees of freedom inside a compound nucleus. In this case it is typical to treat the energies of the resonances (especially if they are individually distinguished) as random numbers with a characteristic mean level spacing $d = 1 / \rho$. Whereas non-interacting energy levels are distributed so that their level spacings obey a Poisson distribution, instead these strongly interacting levels obey a distribution derived by Dyson and Wigner. This distribution is regarded as characteristic of spectra for systems whose classical analogs are chaotic \cite{RevModPhys.69.731,RevModPhys.81.539,Honvault2000233}.

In this energy-domain picture, the resonance widths are related, sometimes in a subtle way, to the Hamiltonian matrix elements $W_{\mu a}$ that couple a bound resonant state $\mu$ of the collision complex, to a scattering state $a$ \cite{peskin:9672,Remacle1996,A808640K}. In the random matrix theory of nuclear scattering, these matrix elements are themselves random numbers, distributed about a mean resonance width ${\bar \Gamma}$.  The theory identifies two distinct regimes of resonant scattering.  In the first, ${\bar \Gamma} / d \ll 1$, meaning that the resonances are resolvable (though still distributed randomly). In the other limit, ${\bar \Gamma } / d \gg 1$, the resonances overlap. Rather than washing out completely, however, the resulting spectrum exhibits ``Ericson fluctuations'' on a scale set by $\bar\Gamma$ itself \cite{PhysRevLett.5.430,Ericson1963390}. Both regimes are observed in nuclear physics, with Ericson fluctuations typically appearing at higher energies \cite{RevModPhys.82.2845}.

In this article we apply the methods of random matrix theory to cold collisions within the Wigner threshold regime.  The object of our study will be atom-diatom collisions, which possess far fewer degrees of freedom than the polyatomic molecules or complex nuclei described above. Nevertheless, it has been well-established that the same ideas apply to nominally ``simpler'' systems, even to the level of a single electron in a diamagnetic Rydberg state \cite{PhysRevLett.65.1100,Pohl2009181} or to conductance fluctuations in a semiconductor device \cite{RevModPhys.69.731}, in the quantum chaos regime.

To apply the statistical model to cold collisions, we must balance the highly resonant, strongly coupled, $10^3$K energy physics of the complex against the delicate sub-mK energy scales of the ultracold. To do this, we exploit ideas of multichannel quantum defect theory (MQDT) \cite{0034-4885-46-2-002,PhysRevLett.81.3355}. This theory makes a clean distinction between the physics of the complex, which is pertinent when the colliding species are close together; and the physics of the long range scattering, which is sensitive to such things as the hyperfine states of the atom and molecule, the low collision energy, and any applied electromagnetic fields. Uniting these two disparate sets of phenomena, we can assess the influence of highly resonant scattering in the threshold regime, and in particular its dependence on the hyperfine state selected for the collision. Although the multichannel scattering cross sections are derived from a fairly realistic framework, we find nevertheless that the simple RRKM rate \eq{RRKM_rate} is a useful tool for interpreting the results, even at ultracold temperature.

The present work is outlined as follows. In Section \ref{sec:theo} we detail our theoretical framework, which is divided into two aspects. Section \ref{sec:theoscattering} introduces the general scattering framework and the treatment of the long-range interactions via MQDT. In Section \ref{sec:statisticalK} we then present our approach of treating the highly resonant short-range part by means of a statistical framework derived from random matrix theory. The essential input parameter for the statistical theory is the density of states for the short-range resonances; in Section \ref{sec:dos} we provide estimates for all non-reactive A + AB alkali dimer pairs. The question of including the density of states due to the nuclear spin degrees of freedom is addressed in Sec.~\ref{sec:nucspins}. In Section \ref{sec:elastic} we present exemplary elastic cross sections within the Wigner threshold law regime that are derived from our theoretical framework. Two particular examples are chosen: K + LiK, where resonances remain well-separated, and Rb + KRb, where the DOS is high. Also, magnetic field dependent thermal rates are provided. Section \ref{sec:ericson} finally shows that ultracold atom-molecule collisions demonstrate the onset of Ericson fluctuations on a completely different energy scale than in nuclear physics. In Section \ref{sec:opportunities} we comment on what might be learned from experimental data by comparing to the predictions and assumptions of our model. With Section \ref{sec:conclusion} we provided a brief conclusion and an outlook on further directions for our theory of highly resonant scattering.


\section{Theoretical framework}
\label{sec:theo}
We deal here with the three-body physics of ultracold alkali atoms, a calculation that could, in principle, be performed in substantial detail \cite{PhysRevA.71.032722,dincao:052709}.  It is, however, an immense labor, and the results, while qualitatively meaningful, are unlikely to be quantitatively accurate.  Even in cases where the calculations can be converged, the relevant potential energy surfaces are not known to sufficiently high accuracy for ultracold collisions.  Nevertheless, the scattering framework is standard. In this section we develop this framework, including our approximate, statistical version of the resonant states.

\subsection{Scattering framework}
\label{sec:theoscattering}
We begin with a diatomic molecule AB (where A and B denote alkali atoms) in its $^1 \Sigma$ electronic ground state, its $v=0$ vibrational ground state, and its $n=0$ rotational ground state, according to the Hund's case b) coupling scheme.  Examples of such molecules have been produced in gases of order $\mu$K temperatures \cite{Ni2008a,deiglmayr:133004,PhysRevLett.105.203001,Danzl2010}.
The molecules may or may not be also prepared in their ground state of nuclear spin $I$ \cite{PhysRevLett.104.030402}. These molecules will collide with another alkali atom C (typically one of A or B), in its $^2 S$ ground state, and with its own accessible hyperfine degrees of freedom.

For the time being, we consider cases where chemical reactions are not energetically allowed at ultralow temperature. Therefore, the only collisions we consider are those that can change the nuclear spin quantum numbers. Generally, we are interested in the regime where the atom's spin state can be labeled by $|f,m_f\rangle$, even in the presence of a magnetic field. For the molecule, we assume a magnetic field sufficiently large that the nuclear spins $I_A$ and $I_B$ are decoupled and the states can be characterized by their individual projections on the magnetic field axis, $|I_AM_{A},I_BM_{B} \rangle$. The observables then consist of the collision rate constants 
\begin{equation}\label{eq:Kobs}
K_{M_A, M_B, fm_f \rightarrow M_A^{\prime}, M_B^{\prime}, f^{\prime}m_f^{\prime}}
= \langle v\sigma_{M_A, M_B, fm_f \rightarrow M_A^{\prime}, M_B^{\prime}, f^{\prime}m_f^{\prime}}\rangle
\end{equation}
where $v$ is the relative velocity before the collision, and $\sigma$ is the collision cross section. We omit the nuclear spins $I_A$ and $I_B$ in \eq{eq:Kobs} since they are not subject to change in the collisions we are considering. Assuming a decomposition into partial waves $|LM_L\rangle$ for the relative motion, the cross section is given by
\begin{align}
&\sigma_{M_A, M_B, fm_f \rightarrow M_A^{\prime}, M_B^{\prime}, f^{\prime}m_f^{\prime}}
= \frac { \pi }{ k^2 }\nonumber\\
&\times\!\!\! \sum_{LM_L L^{\prime}M_L^{\prime}}
\left| 1-S_{M_A^{\prime}M_B^{\prime}f^{\prime}m_f^{\prime}L^{\prime}M_L^{\prime};M_AM_Bfm_fLM_L} \right|^2,
\end{align}
in terms of the scattering matrix elements $S_{a^{\prime}a}$. The $S$-matrix describes the possible re-arrangement of angular momentum during the collision, but must conserve the total projection, $M=M_A^{\prime} + M_B^{\prime} + m_f^{\prime} + M_L^{\prime} + m_n^{\prime} = M_A + M_B + m_f + M_L + m_n$, with the quantization axis applied along the magnetic field direction, if any.  In the above equality we included the projection $m_n$ of the rotational quantum number $n$ of the molecule, which is needed when considering possible resonant states. For the incident and outgoing channels, however, we will always assume the ro-vibrational ground state, i.e., $v=n=m_n=0$. For notational convenience we hereafter denote these scattering channel indices as
\begin{eqnarray}
|a \rangle = |v=n=0,M_A M_B fm_f LM_L \rangle.
\end{eqnarray}
Calculation of a schematic but realistic $S_{a^{\prime}a}$, including its energy- and magnetic field-dependent  resonance structure, is the goal of this article.

\begin{figure}
 \includegraphics[width=8.5cm]{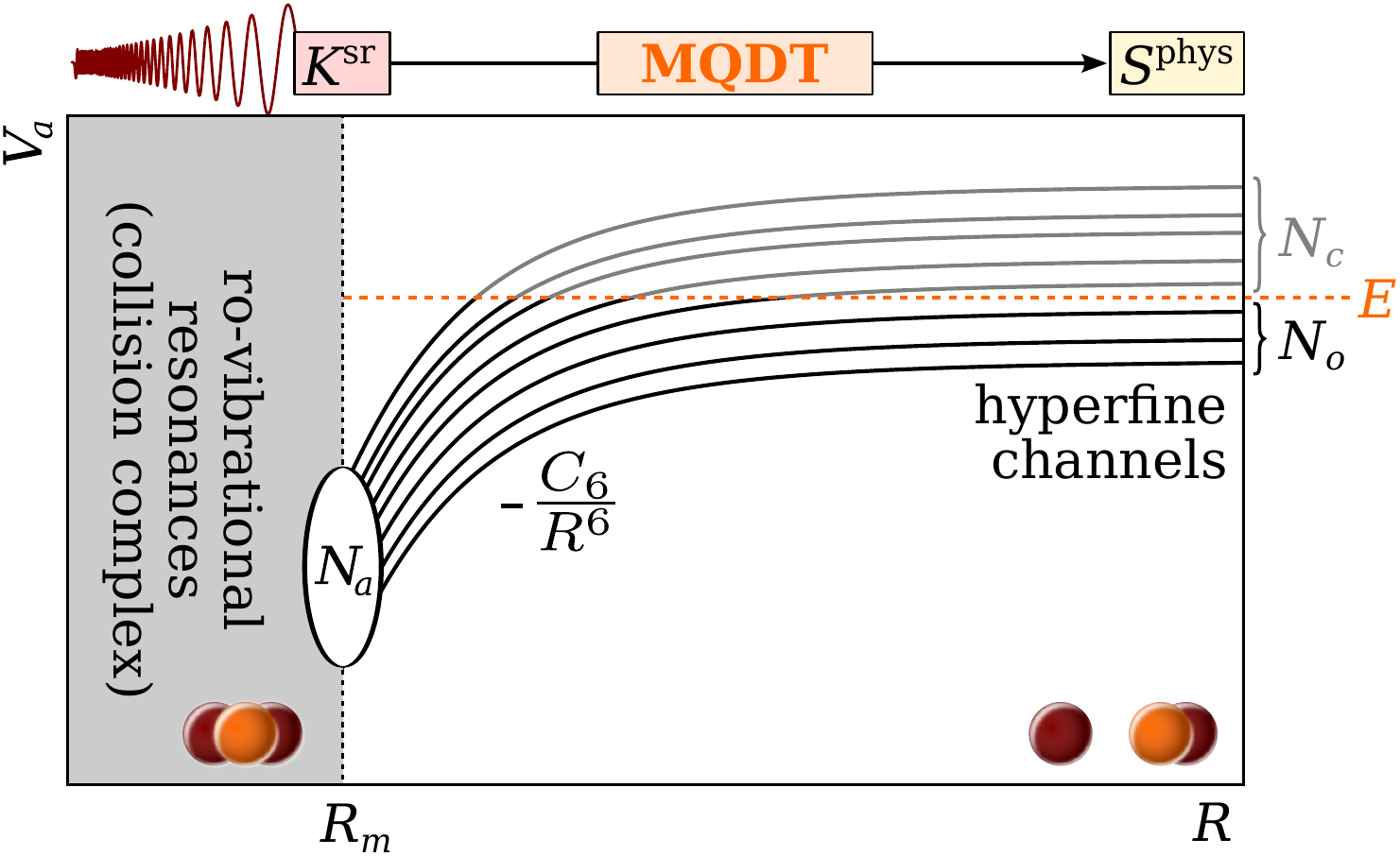}
\caption{(color online) Schematics of our MQDT approach (not to scale). In the long-range part, $R>R_m$, we only consider the ro-vibrational ground state of the molecule, but include all $N_a$ atomic and molecular hyperfine states and different partial waves (not shown). In the asymptotic region, $R\rightarrow\infty$, $N_o$ of them are energetically open and $N_c$ are closed. The MQDT treatment transforms the short-range $K$-matrix $\ksr$, defined at $R_m$, into a physical scattering matrix $\underline{S}^\mathrm{phys}$ from which quantities such as elastic and inelastic cross sections can be deduced. $\ksr$ includes the information on the ro-vibrational resonances which occur in in the short-range part, $R<R_m$.
\label{fig:qdt}}
\end{figure}

To achieve this goal, we exploit the conceptual difference between the spin channels $|a \rangle$ that describe physics at large interparticle separation $R$; and the numerous resonant states $|\mu \rangle$ that differ by rotational and vibrational quantum numbers from $a$, and that describe states of the scattering complex. This general separation of states is illustrated schematically in \fig{fig:qdt}. For separations $R$ greater than some characteristic distance $R_m$, the channels $a$ are assumed to be independent of one another, and described by simplified long-range interactions of the form
\begin{eqnarray}
V_a(R) = - \frac{ C_6 }{ R^6 } + \frac{ \hbar^2 L_a(L_a+1) } {2 m_r R^2}+E_a(B),
\end{eqnarray}
where $E_a(B)$ is the threshold of the $a$th channel, which may depend on a magnetic field $B$. Here, $m_r$ is the reduced mass of the scattering partners and $C_6$ is their van der Waals coefficient, which is taken to be isotropic in this model.

Dividing the scattering process into short- and long-range parts forms the basis of quantum defect theory (QDT). Here, we utilize a multichannel formulation of QDT along the lines of Ref.~\cite{PhysRevLett.81.3355}. The key feature of MQDT is that -- once the MQDT parameters have been determined for a given class of potentials -- one only needs to provide the reactance matrix $\ksr=i(\mathbb{1}-\ssr)(\mathbb{1}+\ssr)^{-1}$ which is defined at the matching radius $R_m$ between the short- and the long-range. The MQDT formalism as outlined in this section then takes care of the propagation for $R>R_m$ and directly yields the physical scattering matrix $\underline S^\text{phys}$, which defines the solution vectors $\boldsymbol{\psi}^{(a)}$ of the coupled channel equations for the whole scattering process,
\begin{align}
 \psi^{(a)}_{a^\prime}(R)\stackrel{R\rightarrow\infty}{=}\delta_{a^\prime a}f^-_a(R)-S^\mathrm{phys}_{a^\prime a}f^+_{a^\prime}(R).
\end{align}
$f^\pm_a=\sqrt{2m_r/\pi\hbar^2 k_a}e^{\pm i(k_aR-L_a\pi/2)}$ are outgoing ($+$) and incoming ($-$) spherical waves, respectively. Once determined, $\underline S^\text{phys}$ can easily be converted into various observables describing the scattering process.

We apply this formalism explicitly only to the small number $N_a$ of hyperfine channels belonging to the ro-vibrational ground state of the molecule and the ground electronic state of the atom. Of these, some number $N_o$ will be energetically open, meaning that for these channels $E>E_a$ and the collision partners can escape to infinity. The remaining $N_c=N_a-N_o$ closed channels do not contribute directly to the physical scattering matrix, and must be ``eliminated'' by the usual algebraic procedures of MQDT.

To do so, the short-range $K$-matrix $\ksr$ is partitioned into its open and closed channels at $R_m$,
\begin{equation}
  \ksr=\begin{pmatrix}
         \ksr_{oo}&\ksr_{oc}\\
	 \ksr_{co}&\ksr_{cc}
       \end{pmatrix}.
\end{equation}
The closed channels are eliminated in the MQDT sense through
\begin{equation}\label{eq:ktilde}
 \ksrtilde=\ksr_{oo}-\ksr_{oc}(\ksr_{cc}+\tan\underline{\beta})^{-1}\ksr_{co},
\end{equation}
where $\beta$ is a closed-channel MQDT parameter \cite{PhysRevLett.81.3355}. The modified reactance matrix $\ksrtilde$ has dimension $N_o\times N_o$ and shows the potential influence of closed channel pathways. The transformation to an energy-normalized, nonanalytic long-range representation is achieved by
\begin{equation}\label{eq:mqdttrans}
 \underline{K}=\underline{A}^\onehalf\ksrtilde(1+\underline{\mathcal G}\ksrtilde)^{-1}\underline{A}^\onehalf.
\end{equation}
The physical scattering matrix is finally formed by
\begin{equation}\label{eq:sphys}
 \underline{S}^\text{phys}=e^{i\underline{\eta}}(1+i\underline{K})(1-i\underline{K})^{-1}e^{i\underline{\eta}}.
\end{equation}

$\underline{A}$, $\underline{\mathcal G}$, $\underline{\eta}$, and $\underline{\beta}$ are diagonal matrices in the asymptotic channel space, consisting of the relevant MQDT parameters. The latter are determined as in Refs.\ \cite{PhysRevLett.81.3355,burkethesis}. In the present from of MQDT we encounter two sets of long-range reference functions: $(f^0,g^0)$ are smooth, analytic functions of energy whereas $(f,g)$ are energy-normalized but non-analytic functions of energy. $(f,g)$ are solutions of the Schr\"odinger equation in presence of a long-range potential $V^\text{lr}(R)$ and are related to the energy-normalized spherical Bessel and Neumann functions via
\begin{equation}\label{eq:defineeta}
 \begin{split}
  f(R)&\xrightarrow{R\rightarrow\infty}kR\sqrt{2m_r/\pi k}\,[j_l(kR)\cos\eta-n_l(kR)\sin\eta]\\
  g(R)&\xrightarrow{R\rightarrow\infty}kR\sqrt{2m_r/\pi k}\,[j_l(kR)\sin\eta+n_l(kR)\cos\eta].
 \end{split}
\end{equation}
Equation (\ref{eq:defineeta}) defines the MQDT parameter $\eta$. The parameter $\beta$ is a negative energy phase that represents the phase accumulated in $V^\text{lr}(R)$. The energy-normalized base pair $(f,g)$ is related to energy-analytic base pair $(f^0,g^0)$ through the transformation
\begin{align}
\label{eq:reffunctrans2}
 \begin{pmatrix}
  f^0\\g^0
 \end{pmatrix}
&{}=
\begin{pmatrix}
 A^{-\frac{1}{2}}&0\\
 -A^{-\frac{1}{2}}\mathcal{G}&A^{\frac{1}{2}}
\end{pmatrix}
 \begin{pmatrix}
  f\\g
 \end{pmatrix},
\end{align}
which defines the MQDT parameters $A$ and $\mathcal G$.

MQDT has been a hugely successful tool for organizing apparently complex spectra of atoms \cite{RevModPhys.68.1015} and in simply describing resonant scattering, both at thermal energies \cite{mies:2514,mies:2526} and in ultracold atom collisions \cite{PhysRevA.62.012708,PhysRevLett.81.3355,PhysRevA.72.042719,gao:012702}. Much of its appeal in these circumstances lies in the fact that a matching radius $R_m$ can be chosen so the channels that will be closed as $R \rightarrow \infty$ remain classically open at $R=R_m$.  If this is so, the short-range $K$ matrix becomes a weakly energy-dependent quantity, and complex spectra can be unified by the simple algebraic procedures described above. For molecular scattering, it remains to be seen whether this same simplicity occurs, since for any $R_m$ there may be many channels that are already classically closed, and hence, impart resonant structure to $\ksr$.  Indeed, there is already some hint in applications of MQDT to cold molecule collisions that $R_m$ must be chosen carefully to maximize the simplicity of $\ksr$ \cite{PhysRevA.84.042703}.

In the present case of highly resonant scattering, we in fact approach quite the opposite limit, where for any reasonable $R_m$ most of the resonant ro-vibrational channels are already closed.  Thus, our $\ksr$ will necessarily be highly energy dependent, exhibiting already the resonances of interest. Although it is difficult to compute, it remains nevertheless a well-defined quantity in the theory.  For our present purposes, we employ MQDT as quick, algebraic solution to producing scattering matrices $\underline{S}^\mathrm{phys}$ for a given $\ksr$. Arriving at a physically reasonable $\ksr$ is the task we turn to next.

\subsection{Statistical short-range $K$-Matrix}
\label{sec:statisticalK}
In treating the long-range collision physics by means of MQDT, the only quantity left to be determined is the short-range $K$-matrix. It is indexed by the $N_a$ asymptotic channels $a$, but is influenced by the myriad (i.e., $N\gg N_a$) of resonant states $\mu$.  Quite generally, it can be expressed as \cite{RevModPhys.82.2845}
\begin{equation}\label{eq:ksr}
 K^\text{sr}_{a,b}(E)=-\pi\sum_{\mu=1}^N\frac{W_{a\mu}W_{\mu b}}{E-E_\mu}.
\end{equation}
Equation (\ref{eq:ksr}) is expressed in the eigenspace of a short-range Hamiltonian $\underline{H}^\text{sr}$ that gives rise to the (unperturbed) short-range levels at $E_\mu$. $W_{a\mu}=W_{\mu a}$ are (assumed-to-be energy independent) coupling matrix elements between resonance $\mu$ and asymptotic channel $a$. The mean coupling strength of the $a$th asymptotic channel to the short-range resonances is given by the dimensionless parameter
\begin{equation}\label{eq:overlap0}
 R_a^{(0)}=\frac{\pi}{2}\rho\bar{\Gamma}_a,
\end{equation}
where $\bar{\Gamma}_a=(2\pi/N)\sum_{\mu=1}^N |W_{\mu a}|^2$ is the zero-order average partial width to the decay channel $a$ \cite{A808640K} and $\rho$ is the DOS of the resonances, evaluated at the incident energy.

The input parameters for the resonant scattering theory, \eq{eq:ksr}, are the zero-order positions $E_\mu$ of the resonances and the coupling elements $W_{a\mu}$ to the asymptotic channels. However, both are usually unknown unless the short-range part is known with high precision. To provide \eq{eq:ksr} with reasonable input parameters, we utilize a statistical model where $E_\mu$ and $W_{a\mu}$ are taken as random variables. This model follows closely the random matrix theory approach in nuclear reaction physics \cite{RevModPhys.82.2845} and can be also found in theoretical works on quantum transport \cite{RevModPhys.69.731}, as well as in the theory of chemical reactions \cite{A808640K}. By employing such a model, we assume that the collision complex corresponds classically to a long, chaotic trajectory that ergodically explores a large portion of the allowed phase space.

Acknowledging the statistical nature of the short-range resonance levels, we apply random matrix theory to $\ksr$ based on the Gaussian Orthogonal Ensemble (GOE) according to \cite{RevModPhys.82.2845}. In particular, we assume that the spectrum $E_\mu$ of the resonant states is determined by a Hamiltonian $H^\text{GOE}$ that is a member of the GOE. As a such, the nearest neighbor distribution of the spectrum satisfies the \emph{Dyson-Wigner distribution}
\begin{equation}\label{eq:goespectrum}
 P(s_\mu)\equiv P(s)=\frac{\pi}{2}se^{\pi s^2/4},
\end{equation}
where $s_\mu=|E_{\mu+1}-E_\mu|/d$ is the nearest-neighbor level spacing in units of the mean level spacing \cite{RevModPhys.81.539}. In practice, we produce the spectrum $E_\mu$ for a given DOS $\rho=1/d$ by constructing first a set $\{s_\mu\}$ of nearest neighbor splittings satisfying \eq{eq:goespectrum} \cite{Hofmann1975391}; the  spectrum is then given by $E_\mu=E_0+\sum_{i=1}^{\mu-1}s_\mu$, where $E_0$ is an appropriately chosen offset. An exemplary GOE spectrum for Rb + KRb scattering is reproduced in \fig{fig:overview} along with its nearest neighbor distribution.

Since $\ksr$ is expressed in the frame where $H^\text{GOE}$ is diagonal, the coupling matrix $\underline{W}$ becomes a random process itself \cite{RevModPhys.82.2845}. More precisely, its elements are given by uncorrelated, Gaussian-distributed random variables with mean 0 and variance $\nu_a^2$. Hence, $\bar\Gamma_a=2\pi\nu_a^2$ are the mean zero-order partial widths. From \eq{eq:overlap0} we find
\begin{equation}\label{eq:xanua}
 \nu_a^2=\frac{R_a^{(0)}}{\rho\pi^2}.
\end{equation}
Thus, in order to describe the short-range physics within the statistical model, it is sufficient to specify the DOS $\rho$ of the short-range resonances and the mean coupling strength $R_a^{(0)}$ to the asymptotic channels. In the present work, we will usually assume $R_a^{(0)}=1$ for which the transmission coefficient $T_a$ between the short and long-range channels \cite{PhysRevLett.5.430,A808640K,RevModPhys.82.2845},
\begin{equation}
 T_a=\frac{4R_a^{(0)}}{\big[1+R_a^{(0)}\big]^2},
\end{equation}
reaches unity. In other words, outbound flux that has left the collision complex and reaches $R_m$, is assumed to continue out with unit probability. Some of this flux will later be reflected back to small $R$ due to details of the hyperfine channels $|a\rangle$. This effect, however, is fully accounted for in MQDT.

Having a unit transmission probability corresponds to the RRKM limit of transition state theory, reached for barrierless reactions. In transition state theory, the decay rate of a metastable state (here: the short-range resonances) is proportional to the ratio between the number $N_a$ of open channels (here: the asymptotic channels) and the level density $\rho$ of the metastable states \cite{levinebook}. Indeed, one recovers for the decay rate $k$,
\begin{equation}\label{eq:kRRKM}
 k=\frac{\sum_{a=1}^{N_a}T_a}{2\pi\hbar\rho}=\frac{N_a}{2\pi\hbar\rho}=k_\mathrm{RRKM}
\end{equation}
when $T_a=1$ \cite{A808640K}. In general, however, $R_a^{(0)}$ can act as a fitting parameter to real spectra, revealing further details on the short-range physics.

\subsection{Ro-vibrational density of states}
\label{sec:dos}

\begin{figure}
\includegraphics[width=8.5cm]{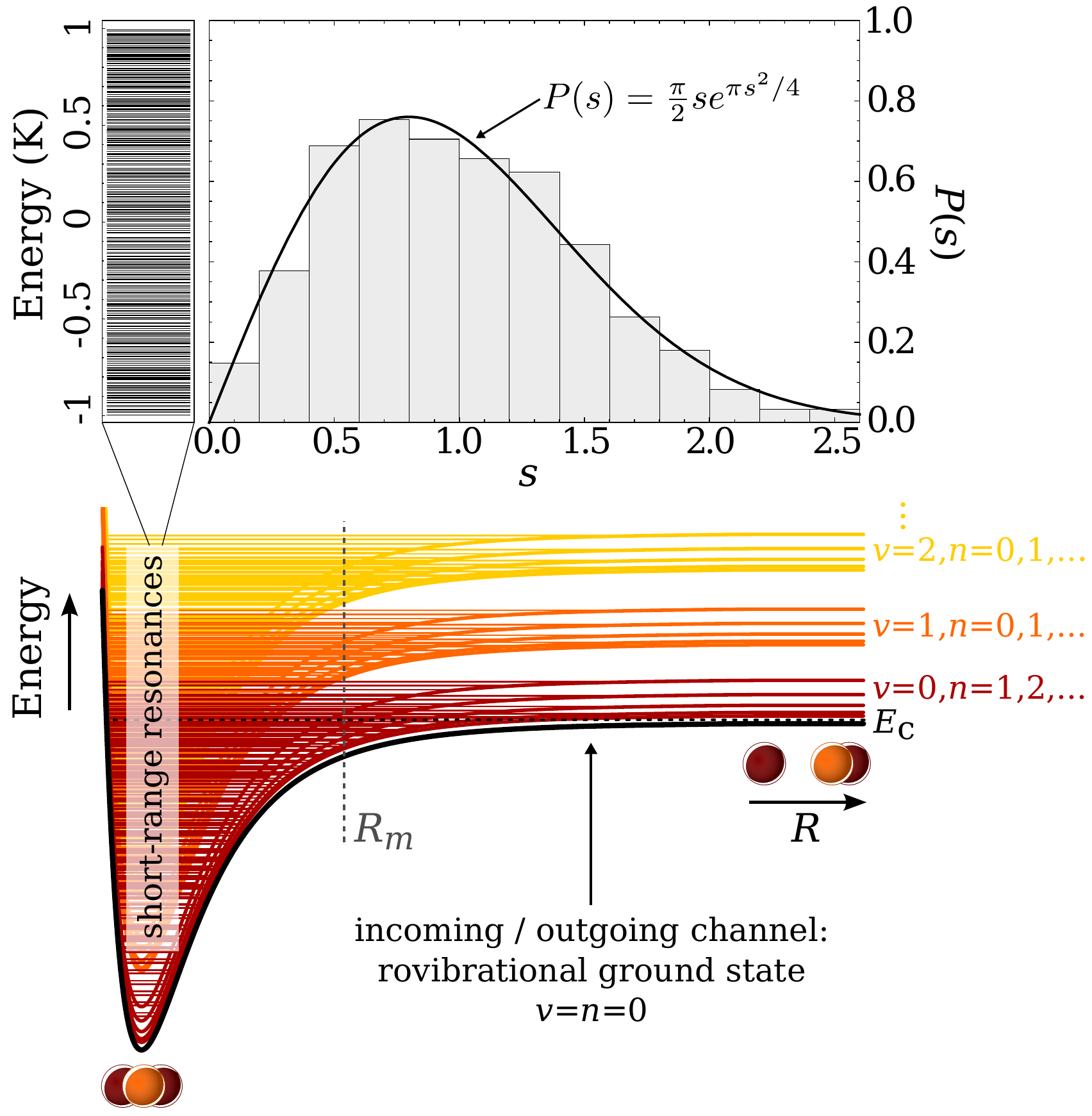}
\caption{(color online) Schematic overview (not to scale) of the origin of the short-range resonances and their distribution. (bottom) The atom-molecule potential is modeled by a Lennard-Jones potential, \eq{eq:LJ}. The resonant channels stem from ro-vibrationally excited states of the molecule. For ultracold temperatures, the incident and outgoing scattering channels are restricted to the ground ro-vibrational state of the molecule. Within this ground state, the various spin states are treated explicitly by means of MQDT. (top) Exemplary short-range spectrum for $s$-wave collisions of Rb + KRb; it is constructed to satisfy the Dyson-Wigner distribution for the nearest neighbors, cf.\ \eq{eq:goespectrum}.
\label{fig:overview}}
\end{figure}

Having the statistical model for the short-range $K$-matrix in hand, \eq{eq:ksr}, the question of computing the mean density of states itself remains. Since in general we do not know the short-range potential in detail, it is impossible to calculate the real atom-diatom ro-vibrational spectrum, which would give rise to the short-range resonances in question. We thus pursue the following strategy to get an adequate estimate of the short-range ro-vibrational density of states $\rho_\mathrm{rv}$ (see also \fig{fig:overview}):
\begin{enumerate}
 \item The short-range interaction is approximated by a Lennard-Jones potential along the reaction coordinate $R$ and add the centrifugal barrier due to the end-over-end rotation angular momentum $L$ of the atom and the molecule about one another,
\begin{equation}\label{eq:LJ}
 V_\text{sr}^{(L)}(R)=\frac{C_{12}}{R^{12}}-\frac{C_6}{R^6}+\frac{L(L+1)}{2m_r R^2}.
\end{equation}
A fairly realistic estimation of the long-range behavior of this potential is to assume $C_{6,\mathrm{B}+\mathrm{AB}}=C_{6,\mathrm{AB}}+C_{6,\mathrm{B_2}}$ \cite{1367-2630-12-7-073041}; a convenient compilation of the $C_6$ coefficients of all alkali dimers can be found in \cite{PhysRevA.82.012510}, and we use these here. The $C_{12}$ coefficient in \eq{eq:LJ} refers to the short-range behavior of the potential. It can be expressed in terms of the overall depth $D_e$ of the potential and the $C_6$ coefficient via $C_{12}=C_6^2/4D_e$; in the present work, we employ the realistic depths $D_e$ as calculated in \cite{PhysRevA.81.060703}. $D_e$ refers to the dissociation energy of the ground state trimer AB$_2$ into AB + B.
 \item For every partial wave $L$ of interest, we calculate the bound state energies $E_\alpha^{(L)}$ of $V_\text{sr}^{(L)}(R)$. $\alpha$ labels the vibrational quantum number in $R$, and each of these states represents a possible short-range resonance.
 \item Each asymptotic ro-vibrational channel ($v,n$) can give rise to such short-range resonance states. With each of these channels we associate a set of resonance energies $E_\alpha^{(L,v,n)}$ that is offset by the corresponding channel threshold, i.e., $E_\alpha^{(L,v,n)}=E_\alpha^{(L)}+E_{v,n}$. The total angular momentum $\mathbf{J=L+n}$ is assumed to be conserved in the usual quantum mechanical way, hence the triangular conditions hold for the possible combinations of $L$ and $n$. In particular, for $J=0$ ($s$-wave collisions), only $L=n$ is possible.
 \item All allowed energies $E_\alpha^{(L,v,n)}$ form a total spectrum. From this spectrum we can extract the mean level spacing $d$ and the level density $\rho_\mathrm{rv}=1/d$ associated with the ro-vibrational resonant states. In doing so, we restrict ourselves to a certain energy interval centered around the incoming channel threshold.
 \item The analysis so far does not account for any degeneracy of the energy levels. We consider the case where only the \emph{total} magnetic quantum number $M$ is conserved. Since $M=M_\mathrm{A}+M_\mathrm{B}+m_f+M_L+m_n$, there are numerous possibilities to couple to a given total $M$. These degeneracies are accounted for when calculating the final DOS.
\end{enumerate}

\begin{table}
\caption{Ro-vibrational DOS $\rho_\mathrm{rv}(\mathrm{mK}^{-1})$ for $^{87}\mathrm{Rb}+{}^{40}\mathrm{K}^{87}\mathrm{Rb}$ collisions as a function of the maximal allowed vibrational (rows) and rotational (columns) quantum numbers. The total angular momentum $\mathbf{J=L+n}$ is assumed to be conserved; the values reported here are for $J=0$, $M_L+m_n=0$. Including more than 25 vibrational and 100 rotational level does not increase the DOS further.}
\begin{ruledtabular}
\begin{tabular}{ccccccc}
$v_\mathrm{max}/n_\mathrm{max}$&10&20&40&60&80&100 \\ \hline
0 &0.12&0.23&0.38&0.53&0.66&0.79\\
2 &0.14&0.29&0.57&0.93&1.28&1.61\\
10&0.17&0.40&0.93&1.65&2.53&3.46\\
20&0.18&0.46&1.20&2.22&3.46&4.81\\
25&0.20&0.49&1.29&2.34&3.59&4.94
\end{tabular}
\end{ruledtabular}
\label{tab:doskrb}
\end{table}

Specific examples of DOS calculated in this way are presented in Tables \ref{tab:doskrb} and \ref{tab:properties}. The former shows the dependence of the resulting DOS on the maximal ro-vibrational quantum numbers $v_\mathrm{max},n_\mathrm{max}$ used in the estimate; the Rb + KRb collision is chosen as a particular example. The DOS increases with the number of allowed ro-vibrational levels $v_\mathrm{max},n_\mathrm{max}$ and saturates if a large number of ro-vibrational levels is included. This saturation is because of the finite sampling interval for the calculation of the mean DOS: the bound states belonging to highly excited ro-vibrational molecular levels simply lie outside the sampling region. We remark that the final densities of states are rather insensitive to the particular sampling interval chosen; in the present work, we use an interval of $\pm5$ K centered at the ro-vibrational ground state. In addition, as can be seen in Table \ref{tab:doskrb}, the dependence of $\rho_\mathrm{rv}$ on $v_\mathrm{max},n_\mathrm{max}$ is rather weak. Thus, for example, if atoms in the collision complex can for some reason only access states up to $n_\mathrm{max}=60$ rather than $n_\mathrm{max}=100$, this would only change our estimate by a factor of two. In general, we expect something like one ro-vibrational resonance per mK for $s$-wave scattering of Rb + KRb.
This estimate assumes that the molecular states of AB are all in their singlet electronic manifold. We estimate that including the (much shallower) triplet states would increase the DOS by $\sim10\%$. Therefore, we do not consider these states in the present work.

\begin{table}
\caption{Properties characterizing the atom-molecule potential, \eq{eq:LJ}, and the resulting ro-vibrational DOS for various atom-molecule pairs. The DOS are calculated for $L=0$. We picked the isotopes $^6$Li, $^{23}$Na, $^{40}$K, $^{87}$Rb, and $^{133}$Cs. The $X\,^1\Sigma^+$ molecular PES needed for the calculation of the ro-vibrational states of the AB molecule are taken from the references provided in the last column.}
\begin{ruledtabular}
\begin{tabular}{lcccccc}
& $D_e$& $C_6$ & $\rho_\mathrm{rv}$ & $E_\mathrm{vdW}$ & $\rho_\mathrm{rv}$ & Refs.\\ 
& (cm$^{-1}$) & (au) & (mK$^{-1}$) & (mK) & ($E_\mathrm{vdW}^{-1}$)\\
\hline
Na+LiNa & 2872 & 3015  & 0.19 & 0.569  & 0.11 & \cite{Marques2008} \\ 
K+LiK   & 2124 & 6190  & 0.33 & 0.184  & 0.06 & \cite{PhysRevA.79.042716} \\ 
Rb+LiRb & 2036 & 7159  & 0.76 & 0.056  & 0.04 & \cite{10.1063/1.3524312} \\ 
Cs+LiCs & 2502 & 9785  & 1.41 & 0.026  & 0.04 & \cite{staanum:042513} \\ 
K+NaK   & 1851 & 6297  & 1.22 & 0.150  & 0.18 & \cite{Gerdes2008} \\ 
Rb+NaRb & 1752 & 7273  & 2.73 & 0.050  & 0.14 & \cite{pashov:062505} \\ 
Cs+NaCs & 2186 & 9910  & 4.96 & 0.024  & 0.12 & \cite{O2006} \\ 
Rb+KRb  & 1684 & 8798  & 4.95 & 0.041  & 0.20 & \cite{pashov:022511} \\ 
Cs+KCs  & 1821 & 11752 & 7.68 & 0.020  & 0.16 & \cite{ferber:244316} \\ 
Cs+RbCs & 1825 & 12135 & 12.11& 0.017  & 0.21 & \cite{PhysRevA.83.052519} 
\end{tabular}
\end{ruledtabular}
\label{tab:properties}
\end{table}

We provide similar estimates of ro-vibrational DOS for various collision partners in Table \ref{tab:properties}, assuming that all energetically allowed $v$ and $n$ states contribute. We also include in this table the basic molecular data from which the DOS estimates were obtained. As one might expect, $\rho_\mathrm{rv}$ is larger for heavier collision systems. In particular, the DOS for Cs + RbCs collisions is two orders of magnitude higher than for the light Na + LiNa collision complex. A useful way to express $\rho_\mathrm{rv}$ is in units of states per van der Waals energy $E_\mathrm{vdW}=\hbar^3 (2m_r)^{-3/2}C_6^{-1/2}$, see also Table \ref{tab:properties}. In this representation, the larger van der Waals energy scale of lighter molecules compensates for their smaller number of bound states. In the end, all considered DOS are roughly the same, namely, $\rho_\mathrm{rv}(E_\mathrm{vdW}^{-1})\approx0.1$ within a factor of 3. 

The DOS provided in Tables \ref{tab:doskrb} and \ref{tab:properties} are specific examples for $J=0$, for which $L=n$ needs to be satisfied in order to conserve the total angular momentum $\mathbf{J=L+n}$. For $J\ne0$, there are $2J+1$ possibilities for $\mathbf L$ and $\mathbf n$ to couple ($L=n$, $L=n\pm1$, \dots, $L=n\pm J$) and therefore the DOS increases by approximately the same factor.

We remark that in the above considerations the presence of nuclear spin degrees of freedom is not taken into account. In the following subsection we will therefore discuss the possible influence of the spin on the densities of states.

\subsection{The role of nuclear spins}
\label{sec:nucspins}
Thus far we have considered only the density of states due to rotations and vibrations, i.e., due to the relative motion of the three alkali atoms, denoted by $\rho_\mathrm{rv}$. The DOS will multiply, however, if the nuclear spin degrees of freedom become involved. To see whether the nuclear spin may change during the collision, we employ a semiclassical analysis as follows.  Once the collision complex is formed, it lives, on average, for an amount of time $\tau$ that is related to the mean resonance width by $\tau = \hbar / {\bar \Gamma}$.  During this time, the nucleus of any given atom follows a chaotic trajectory through phase space, according to our ergodic assumption. 
The nuclear spins are influenced during this time by a hyperfine Hamiltonian $H_\mathrm{hf}$ that varies in time as the collision complex explores the phase space. The dominant part of this Hamiltonian arises from the magnetic dipole interaction of the nuclear spin with the spin of the electron immediately in orbit above it in the same atom. The electron spin, however, is subject to fluctuations during this classical trajectory. We therefore expect that the nuclear spin experiences a rapidly time-varying change in its Hamiltonian, $\delta H_\mathrm{hf}(t)$.

The magnitude of these fluctuations can be estimated by the following  argument. We regard the collision complex semiclassically as a repeated set of mini-collisions, occurring at average time intervals $\Delta t$. For instance, at one moment the complex might resemble $\mathrm{A} + (\mathrm{BC})^*$, i.e., the A atom is loosely bound to a molecule BC, which is excited into some ro-vibrational state. Because BC is excited, A cannot escape, but rather returns to collide again.  This collision might result in a different complex, say $\mathrm{B}+(\mathrm{AC})^*$. Consider then the nucleus attached to atom A. Before this collision, this nucleus experiences the unpaired electron on the $A$ atom, and hence essentially the entire hyperfine interaction determined by the corresponding magnetic dipole constant $A_\mathrm{hf}$ \cite{RevModPhys.49.31}. After the collision, the atom A is locked into a singlet state with atom C, and the nucleus sees no hyperfine interaction at all, apart from the modest nuclear quadrupole interaction. Thus, the hyperfine interaction experienced by any given nucleus in the complex is effectively switched randomly between full strength and zero, at random intervals $ \sim \Delta t$. 

Let $\hbar  \omega_{12}$ denote the energy difference between two nuclear spin states in the absence of these fluctuations. Then a nucleus initially in one of the states will end up in the other at time $\tau$ with a probability amplitude
\begin{eqnarray}
\label{spin_perturbation}
c(\tau) = \frac{ 1 }{i \hbar } \int_0^{\tau} \mathrm{d}t\, e^{ i \omega_{12} t}
\delta H_\mathrm{hf}(t)=\frac{\sqrt{2\pi}}{i\hbar}\delta {\tilde H}_\mathrm{hf}(\omega_{12})
\end{eqnarray}
in terms of the Fourier transform $\delta {\tilde H}_\mathrm{hf}(\omega_{12})$.
The perturbing Hamiltonian $\delta H_\mathrm{hf}(t)$ will fluctuate on a characteristic time scale $\Delta t$, set roughly by the mean collision time of an atom in the complex with another atom. The Fourier transform of $\delta H_\mathrm{hf}(t)$ is then nonzero only over some finite bandwidth $\Omega = 2\pi/\Delta t$.  To conserve the intensity of the fluctuations over $\tau$ in the time domain and $\Omega$ in the frequency domain, the root-mean-squared averages of the fluctuation and its Fourier transform must satisfy
\begin{eqnarray}
\sqrt{ \langle [ \delta H_\mathrm{hf}(t)]^2 \rangle_t } \sqrt{\tau} \approx
\sqrt{ \langle [ \delta {\tilde H}_\mathrm{hf}(\omega)]^2 \rangle }_\omega \sqrt{\Omega}.
\end{eqnarray}
If we assume that the time domain fluctuations are random white noise, then the power spectrum is approximately independent of frequency within the bandwidth $\Omega$, i.e., $\delta {\tilde H}_\mathrm{hf}(\omega)=\delta {\tilde H}_\mathrm{hf}^{(0)}=\mathrm{const.}$ and $\sqrt{ \langle [ \delta {\tilde H}_\mathrm{hf}(\omega)]^2 \rangle }_\omega=\delta {\tilde H}_\mathrm{hf}^{(0)}$ correspondingly.  In particular, we assign it this value at the transition frequency, $\delta {\tilde H}_\mathrm{hf}(\omega_{12})=\delta {\tilde H}_\mathrm{hf}^{(0)}$. Employing $c(\tau)=\sqrt{2\pi}/i\hbar\times\delta {\tilde H}_\mathrm{hf}(\omega_{12})$ and $\sqrt{ \langle [ \delta H_\mathrm{hf}(t)]^2 \rangle_t }=A_\mathrm{hf}/2$, we find for an estimate of the transition probability
\begin{align}
\label{transition_probability}
P = |c( \tau )|^2 &=
\left| \frac{ \sqrt{2\pi} }{ i \hbar }  \sqrt{ \langle [ \delta H_\mathrm{hf}(t)]^2 \rangle_t }
\sqrt{ \frac{ \tau }{ \Omega } } 
\right|^2 \nonumber \\
& =\pi^2
\left( \frac{ A_\mathrm{hf} \tau }{ h} \right)
\left( \frac{ A_\mathrm{hf} \Delta t }{ h } \right).  
\end{align}
The first factor in parentheses denotes the size of the perturbation, times the length of time is acts, which would be the probability that a smoothly-varying perturbation changes the spin state. The second factor in parentheses accounts for the fluctuations on a time scale $\Delta t$. Once the perturbing Hamiltonian takes a  certain value, the nuclear spin has only a time $\Delta t$ to respond to this perturbation, i.e., by precessing around the instantaneous local magnetic field. After time $\Delta t$, the perturbation randomly switches to something else, and the nuclear spin attempts to follow a new local field. If $\Delta t$ is much smaller than the nuclear-spin-changing period $\hbar / \omega_{12}$ (as it is in our case), then the nuclear spin has a hard time changing at all; more rapid collisions actually reduce the transition probability. On the other hand, if the collisions occur rarely on the time scale $\hbar / \omega_{12}$, then the relevant $\Delta t$ is reciprocal to the hyperfine interaction itself. In this case, the second factor in (\ref{transition_probability}) is unity, and we reduce to the familiar case of a slowly-varying perturbation.  

\begin{table}
\caption{Mean collision time $\Delta t$ [\eq{eq:meancolltime}], lifetime $\tau$, nuclear spin transition probability $P$ [\eq{transition_probability}], and nuclear spin enhancement factor $N_\mathrm{nuc}$ [\eq{eq:Nnuc}] for various collision complexes in case of $L=0$.}
\begin{ruledtabular}
\begin{tabular}{lcccc}
& $\Delta t$\,(s) & $\tau$\,(s) & $P$ & $N_\mathrm{nuc}$\\ 
\hline
Na+LiNa & $9.1\times10^{-12}$ & $9.4\times10^{-9}$ & $6.6\times10^{-1}$ & 4 \\ 
K+LiK   & $1.7\times10^{-11}$ & $1.6\times10^{-8}$ & $2.1\times10^{-1}$ & 9 \\ 
Rb+LiRb & $2.8\times10^{-11}$ & $3.6\times10^{-8}$ & $1.2\times10^{2}$  & 4 \\ 
Cs+LiCs & $3.9\times10^{-11}$ & $6.8\times10^{-8}$ & $4.3\times10^{2}$  & 4 \\ 
K+NaK   & $3.7\times10^{-11}$ & $5.8\times10^{-8}$ &               1.7  & 16 \\ 
Rb+NaRb & $6.4\times10^{-11}$ & $1.3\times10^{-7}$ & $9.7\times10^{2}$  & 4 \\ 
Cs+NaCs & $9.1\times10^{-11}$ & $2.4\times10^{-7}$ & $1.1\times10^{3}$  & 4 \\ 
Rb+KRb  & $1.0\times10^{-10}$ & $2.4\times10^{-7}$ & $2.8\times10^{3}$  & 31 \\ 
Cs+KCs  & $1.5\times10^{-10}$ & $3.7\times10^{-7}$ & $2.8\times10^{3}$  & 91 \\ 
Cs+RbCs & $2.3\times10^{-10}$ & $5.8\times10^{-7}$ & $6.9\times10^{3}$  & 4
\end{tabular}
\end{ruledtabular}
\label{tab:properties2}
\end{table}

Representative values of $P$ in case of $s$-wave collisions are given in Table \ref{tab:properties2} for some of the atom-molecule pairs we are considering. For calculating these values, we estimate the complex lifetime by means of its RRKM value, $\tau=\hbar/\bar\Gamma=2\pi\hbar\rho_\mathrm{rv}$, where $\rho_\mathrm{rv}$ is the ro-vibrational DOS of the collision complex as provided in Table \ref{tab:properties} for $L=0$. 

The mean collision time can be estimated by averaging over classical trajectories in a pure $C_6$ potential as follows. The time the atom needs to get out of the collision complex, climbing the potential $V_a(R)$ until the classical turning point $R_0$ where the kinetic energy vanishes, and then falling back into the complex, can be approximated by twice the time it needs to fall from $R_0$ all the way in,
\begin{align}
\Delta t_0(R_0) &=2 \int_0^{R_0} \frac{ dR }{ v(R) } = \sqrt{ \frac{2 m_r }{ C_6 } }
R_0^4 \int_0^1 dx \left( \frac{ 1 }{ x^6 } -1\right)^{-1/2} \nonumber\\
&= \frac{\Gamma(\frac{2}{3})}{\Gamma(\frac{1}{6})}
\sqrt{ \frac{2\pi m_r }{ C_6 }} R_0^4, 
\label{eq:colltime}
\end{align}
where $v(R)=\sqrt{2[E-V_a(R)]/m_r}=R_0^3\sqrt{2C_6/m_r}\sqrt{R_0^6/R^6-1}$
is the atom's classical velocity as a function of $R$. The mean collision time $\Delta t$ follows from averaging over all turning points, starting at the equilibrium position $R_e$ of the potential up to some maximal $R_\mathrm{max}$,
\begin{align}
 \Delta t&=(R_\mathrm{max}-R_e)^{-1}\int_{R_e}^{R_\mathrm{max}}dR_0\Delta t_0(R_0)\nonumber\\
&=\frac{1}{5}\frac{\Gamma(\frac{2}{3})}{\Gamma(\frac{1}{6})}\sqrt{\frac{2\pi m_r}{C_6}}\left(1-R_e/R_\mathrm{max}\right)^{-1}R_\mathrm{max}^4.
\label{eq:meancolltime}
\end{align}
The outermost turning point considered, $R_\mathrm{max}=(C_6/2B_\mathrm{rot}^\mathrm{dimer})^{1/6}$, is reached if the collision complex scatters into the energetically lowest closed channel, which is the first rotationally excited state of the molecule with $E=-2B_\mathrm{rot}^\mathrm{dimer}$. Explicit values of $\Delta t$ for various molecules cover the nanosecond to picosecond regime, cf.\ Table \ref{tab:properties2}. Using a pure $C_6$ potential as in \eq{eq:colltime} instead of our model Lennard-Jones potential and integrating all the way to zero instead of stopping at $R_e$ is an excellent approximation; in the case of Rb + KRb, for example, the introduced error is less than 1 \textperthousand.

In cases where $P$ is of order unity or larger (typical for heavy molecules), the nuclear spin is almost certain to change during the lifetime of the complex. Therefore, the nuclear spin degree of freedom also contributes to the total DOS. For constructing our statistical models we would then use
\begin{eqnarray}\label{eq:rhohf}
\rho = \rho_\mathrm{rv}\times N_\mathrm{nuc},
\end{eqnarray}
where $N_\mathrm{nuc}$ denotes the number of nuclear spin states. The latter is determined via
\begin{eqnarray}\label{eq:Nnuc}
N_\mathrm{nuc} = \sum_{M_L=-L}^L f(M,M_L),
\end{eqnarray}
where $f(M,M_L)$ is the number of possible spin states $|M_AM_Bfm_f\rangle$ that conserve the total magnetic quantum number $M$ for a given $M_L$. We remark that in principle the ro-vibrational DOS $\rho_\mathrm{rv}$ also depends on $M_L$; however, this dependence is negligible such that \eq{eq:rhohf} is a valid approximation. Specific examples of $N_\mathrm{nuc}$ are listed in Table \ref{tab:properties2}. In calculating $N_\mathrm{nuc}$ we include all hyperfine states of the atom but restrict ourselves to singlet molecular states. Triplet molecular states lead to excited electronic quartet and doublet states of the triatomic collision complex which are connected to the considered doublet ground state via avoided crossings \cite{Simoni2006}. The contribution of these states can be assessed by calculating their DOS at the threshold of the ground state doublet potential. Our estimate shows that the overall DOS would increase only on the order of 10\%. Hence, we continue to focus on singlet molecular states solely.

Our estimate of nuclear spin-changing probability $P$ is admittedly crude, and represents at best an order-of-magnitude estimate. Nevertheless, all we really need to know is whether $P$ is likely to be much larger than unity. For heavier molecules, this appears to be the case, and we will include nuclear spins in the DOS for our Rb + KRb example below. However, for some lighter species, such as K + LiK, we expect nuclear spins to be fairly well conserved during the collision.

The determination of the complex lifetime $\tau$ is not influenced by the inclusion of the nuclear spin states since both the DOS as well as the number of asymptotic channels increase by the same factor: $\tau=2\pi\hbar\rho/N_a=2\pi\hbar(\rho_\mathrm{rv}\times N_\mathrm{nuc})/N_\mathrm{nuc}=\tau=2\pi\hbar\rho_\mathrm{rv}$. 


\section{Highly resonant scattering near threshold}
\label{sec:results}
Using the model described above, we now calculate simulated collision cross sections in ultralow energy limit.

\begin{figure}
\includegraphics[width=8.5cm]{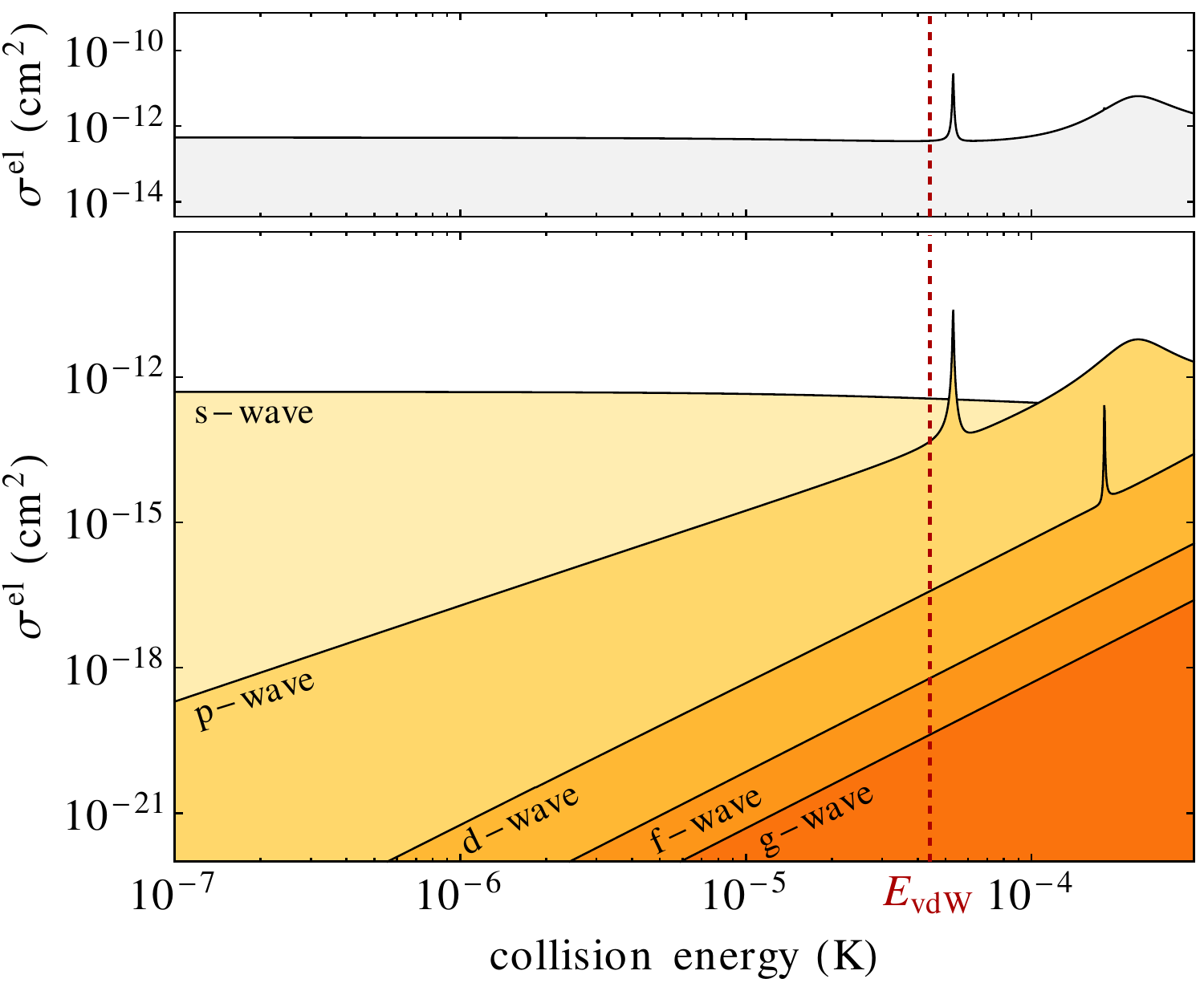}
\caption{(color online) Elastic partial wave cross sections for K + LiK collisions; the top panel shows the corresponding cumulative elastic cross sections. The incident channel is the absolute ground state; the maximum collision energy is twice the van der Waals energy scale $E_\mathrm{vdW}$.
\label{fig:csklik}}
\end{figure}

\subsection{Elastic scattering}
\label{sec:elastic}
We have seen that the density of states can vary widely, depending on the particular species we consider. For this reason, in this section we will explore two schematic cases, where the DOS is either ``low'' or ``high,'' meaning few or many resonant states within the characteristic energy scale $E_\mathrm{vdW}$ within which the Wigner threshold laws hold. 

For the weakly-resonant case, we pick $^{40}$K+$^6$Li$^{40}$K. For this particular example, we do not expect nuclear spin states to be changed ($P=0.2$) and hence expect only $0.06$ $s$-wave resonances per $E_\mathrm{vdw}$, cf.\ Tables \ref{tab:properties} and \ref{tab:properties2}. Exemplary elastic partial wave cross sections up to $L=4$ as a function of the collision energy are shown in \fig{fig:csklik}; the incident channel is the absolute ground state of both the atom and the molecule, so only elastic scattering is possible. As expected from the low DOS, within the given energy range (up to $2\times E_\mathrm{vdW}$) resonances are encountered only sporadically and are thus well resolved. Therefore, the Wigner-law behavior of the elastic cross sections is evident: $\sigma^\mathrm{el}\propto E^{2L}$ for $L=0,1$, and $\sigma^\mathrm{el}\propto E^3$ for $L\ge2$ \cite{Sadeghpour2000}. We remark that the resonances found in \fig{fig:csklik} are determined within our statistical approach and hence are representative, not predictive. For a \emph{quantitative} description of low-resonant cases such as K + LiK a full coupled channel calculation is necessary, at least to provide a realistic short-range $K$-matrix; the long-range part then may still be treated by means of MQDT \cite{PhysRevA.84.042703}.

\begin{figure}
\includegraphics[width=8.5cm]{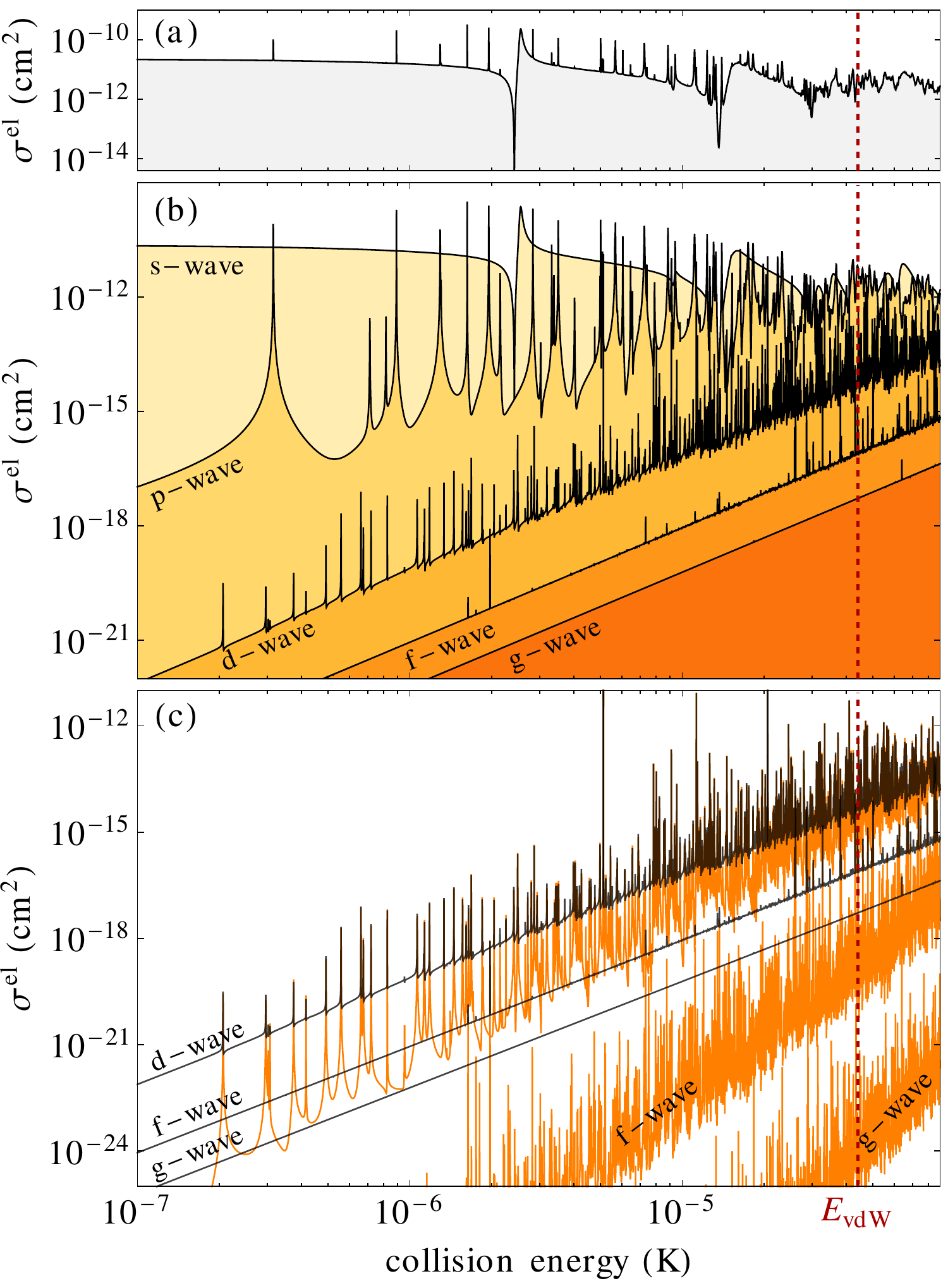}
\caption{(color online) Elastic cross section for Rb + KRb collisions in the absolute ground state. The maximum collision energy is twice the van der Waals energy scale $E_\mathrm{vdW}$. Panel (a) shows the sum of the individual partial waves depicted in (b). Panel (c) provides a comparison of higher ($L\ge2$) partial wave cross sections including [black lines, same as in (b)] and omitting (orange/gray lines) the long-range phase shift $\tan\delta\propto k^4$ due to the $C_6/R^6$ van der Waals dispersion potential.
\label{fig:csrbkrb}}
\end{figure}

Let us now switch to a high DOS, for which the present theory is intended. As a particular example we choose $^{87}\mathrm{Rb}+^{40}\mathrm{K}^{87}\mathrm{Rb}$ collisions \cite{Ospelkaus2010} for which we expect the nuclear spin states to be changed during the formation of the collision complex ($P>1$). Because of Eqs.~(\ref{eq:rhohf},\ref{eq:Nnuc}), the actual DOS depends on the partial wave considered. For higher partial waves $L$, the $2L+1$ projections $M_L$ of the orbital angular momentum allow for a greater variety of spin and rotational states that conserve a given total magnetic quantum number $M$. The resulting DOS for Rb + KRb as a function of $L,M_L$ are tabulated in Table \ref{tab:doskrbpartialwave}. Not only does the DOS increase rapidly as a function of $L$, but also all angular momentum projections $M_L$ need to be summed to form the final partial wave cross section. Since to every $M_L$ a different short-range spectrum is attached, this increases the DOS by an additional factor of approximately $2L+1$ compared to the case of a single $M_L$. This rapid increase of the number of resonances can be observed in \fig{fig:csrbkrb}(b). As indicated by Table \ref{tab:doskrbpartialwave}, within one $E_\mathrm{vdW}$ there are fewer than 10 resonances for $s$-wave collisions, over one hundred for $p$-wave, and already close to one thousand for $d$-wave. 

\begin{table}
\caption{Density of states $\rho$ for $^{87}\mathrm{Rb}+^{40}\mathrm{K}^{87}\mathrm{Rb}$ collisions as a function of the partial wave and its magnetic quantum number. The total magnetic quantum number is always chosen such that the absolute ground state is included.
\label{tab:doskrbpartialwave}}
\begin{ruledtabular}
\begin{tabular}{rrrr@{\hspace{1cm}}rrrr}
$L$&$M_L$&$\rho$\,(G$^{-1}$)&$\rho$\,($\mu$K$^{-1}$)&$L$&$M_L$&$\rho$\,(G$^{-1}$)&$\rho$\,($\mu$K$^{-1}$)\\
\hline
 0& 0 & 5.1   &0.15 &3& 1 & 226.9 & 6.76 \\
 1& -1& 39.9  &1.19 &3& 2 & 226.9 & 6.76 \\
 1& 0 & 44.2  &1.32 &3& 3 & 213.7 & 6.36 \\
 1& 1 & 46.2  &1.37 &4& -4& 157.8 & 4.70 \\
 2& -2& 85.9  &2.56 &4& -3& 203.1 & 6.05 \\
 2& -1& 104.3 &3.11 &4& -2& 248.5 & 7.40 \\
 2& 0 & 117.2 &3.49 &4& -1& 291.0 & 8.67  \\
 2& 1 & 123.6 &3.68 &4& 0 & 325.0 & 9.68 \\
 2& 2 & 123.6 &3.68 &4& 1 & 344.7 & 10.26  \\
 3& -3& 123.9 &3.69 &4& 2 & 344.7 & 10.26 \\
 3& -2& 158.3 &4.71 &4& 3 & 325.0 & 9.67 \\
 3& -1& 189.3 &5.64 &4& 4 & 291.0 & 8.67 \\
 3& 0 & 213.7 &6.36
\end{tabular}
\end{ruledtabular}
\end{table}

For ultracold temperatures usually all collision processes except for $s$-wave are suppressed. In the case of highly resonant scattering as we investigate here, however, there are plenty of resonance peaks due to higher partial waves, cf.\ \fig{fig:csrbkrb}(a). Due to the threshold scaling of resonance widths \cite{PhysRevA.60.414}, these events are isolated and should in principle be well resolvable at the very cold end, $E_c\ll E_\mathrm{vdW}$. Closer to $E_\mathrm{vdW}$, on the other hand, higher partial waves are not as suppressed and in addition resonances start to overlap. Hence, in this regime the appearance of the total cross section eventually is no longer determined by the background scattering cross section with a few resonances on top of it, but rather by the interplay of many overlapping resonances.

The increasing number of resonant states with increasing partial wave is not the whole story, however. As \fig{fig:csrbkrb}(b) shows, the number of visible resonances increases from $L=0$ to $1$ to $2$, but fewer resonances appear for $L>2$. The reason for this can again be found in the Wigner threshold laws. Recall that the elastic scattering phase shifts for a van der Waals potential have two distinct components. There is a short-range component $\delta_\mathrm{sr} \propto k^{2L+1}$, which vanishes faster with energy for higher partial waves because the incident wave function has an ever-greater centrifugal barrier to tunnel through. There is also a long-range component $\delta_\mathrm{lr} \propto k^4$, which arises from scattering outside the outer classical turning point of the centrifugal potential \cite{Sadeghpour2000}. Since the resonances originate in short-range scattering, they appear in $\delta_\mathrm{sr}$, whereby this part of the cross section can be dwarfed by $\delta_\mathrm{lr}$ for $L>2$. To show this more explicitly, we separate out the short-range contribution in \fig{fig:csrbkrb}(c); within the MQDT theory, this amounts to neglecting the $e^{i\eta}$ terms in \eq{eq:sphys}. We therefore conclude that, while the number of resonances grows rapidly with increasing partial waves, nevertheless they are unlikely to be observed in the ultracold.

\begin{figure}
\includegraphics[width=8.5cm]{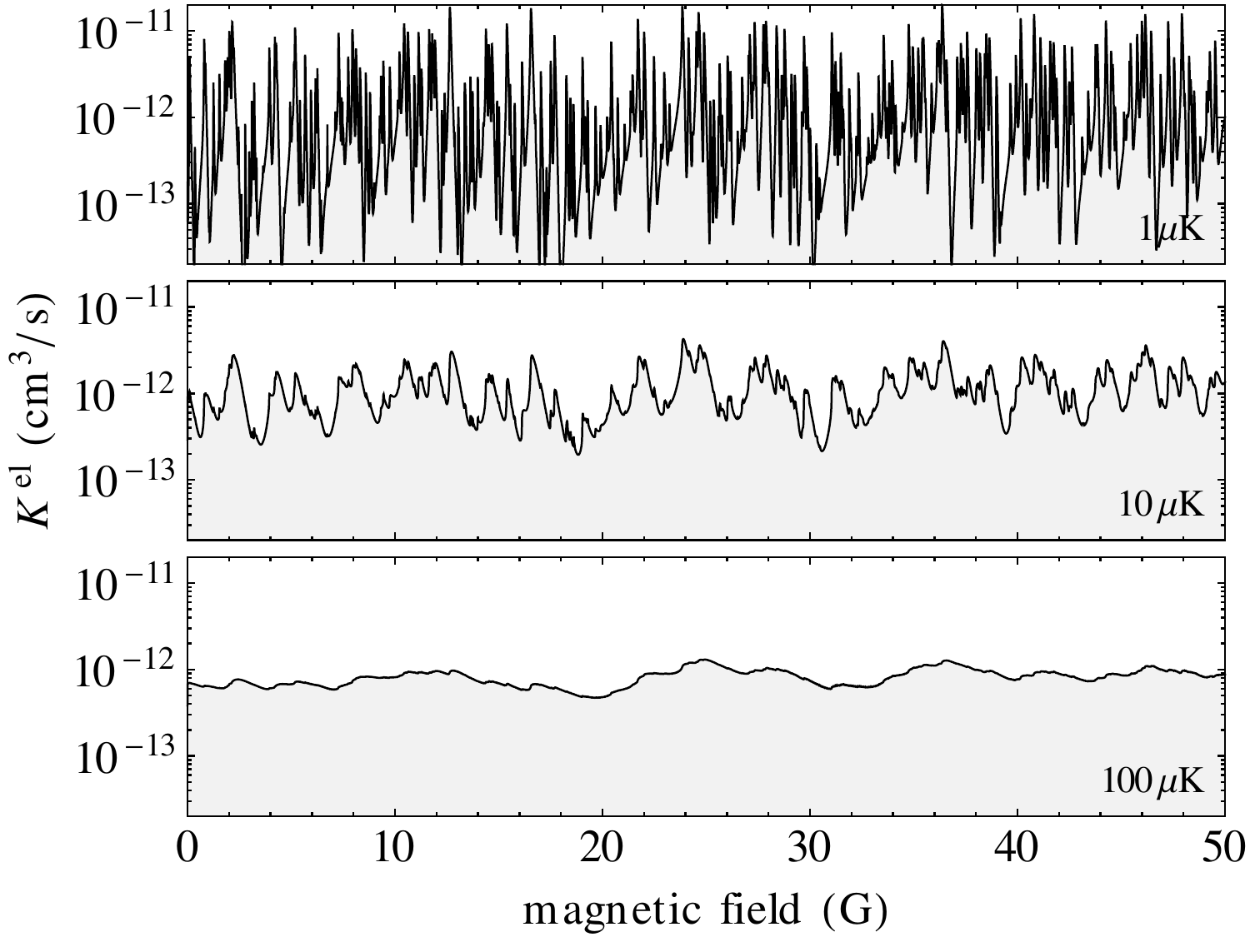}
\caption{Elastic $s$-wave rate constants for Rb + KRb collisions in the absolute ground state. Depicted is the thermalized rate constant for a temperature of 1 $\mu$K, 10 $\mu$K, and 100 $\mu$K, respectively (top to bottom).
\label{fig:csbfield}}
\end{figure}

Instead of the elastic cross sections as a function of collision energy, presented in Fig.~\ref{fig:csrbkrb}, in experimental practice one is more likely to measure scattering rate constants that are thermally averaged. Moreover, often the temperature is fixed and, instead, an external magnetic field is varied to tune the various scattering channels with respect to each other. For these reasons, we provide in \fig{fig:csbfield} the thermally averaged elastic rate constant $K^\mathrm{el}=\langle v\, \sigma^\mathrm{el}\rangle$ as a function of magnetic field, again for Rb + KRb. Because of the vast difference in energy scales, we assume that the short-range physics is independent of the applied magnetic field. As a consequence, the collision complex probes the short-range resonance spectrum with a rate corresponding to the Zeeman shift of the incident channel, i.e., energies are converted into magnetic field strengths via $E=-\mu_\mathrm{mag}B$. For the particular example of $^{87}\mathrm{Rb}+{}^{40}\mathrm{K}^{87}\mathrm{Rb}$ with the absolute ground state as incident channel, this Zeeman shift is largely determined by the magnetic moment of the $f=m_f=1$ ground state of the rubidium atom ($\mu_\mathrm{mag}=0.7$ MHz/G). The resulting densities of states as a function of magnetic field are listed in Table \ref{tab:doskrbpartialwave}; for $s$-wave collisions as depicted in Fig.~\ref{fig:csbfield}, it amounts to 5 resonances per Gauss, which still should be experimentally resolvable.

These resonances will naturally wash out with increasing temperature. In Figure \ref{fig:csbfield} we compare the $s$-wave elastic rate constant for three different temperatures, namely, 1 $\mu$K, 10 $\mu$K, and 100 $\mu$K. As expected, a higher temperature gradually smoothes the sharp resonance peaks found for 1 $\mu$K. Moreover, the few resonances found for $s$-wave scattering as a function of collision energy, \fig{fig:csrbkrb}, have now turned into a dense spectrum of resonances due to the large Zeeman shift of the rubidium atom. Since sub-$\mu$K temperatures are easily reached in ultracold alkali experiments, we predict that individual resonances ought to be observable.

We remark that in principle \fig{fig:csbfield} shows also Fano-Feshbach resonances that occur within the ro-vibrational ground state, i.e., without the need of a highly resonant short-range part. However, within the magnetic field range shown in \fig{fig:csbfield}, there are only a few such resonances: the difference between the last and last but one vibrational level of Rb + KRb is on the order of 100 mK. The atomic Zeeman shift is $33.5\,\mu$K/G, i.e., within 100 G it is unlikely to find two different vibrational states of the same channel.

\subsection{Onset of Ericson fluctuations}
\label{sec:ericson}
Resolving individual scattering resonances requires that their mean width $\bar\Gamma$ be less than their mean separation $d=1/\rho$. However, as more open channels become available, the widths should increase, and when $\bar\Gamma>d$ the scattering should be in the Ericson regime. For ultracold collisions, adding more open channels is as simple as preparing the molecules in a higher-energy hyperfine state. In this section we therefore explore resonance widths as a function of the number $N_o$ of open channels.

In the limit where the mean resonance width exceeds the mean level spacing one might expect the spectrum to become increasingly smooth and featureless. Ericson showed that surprisingly this is not so \cite{PhysRevLett.5.430,Ericson1963390}. Rather, there remains structure that can be probed via the 2-point correlation function $F(\Delta B)$
 (here written in terms of magnetic field strength),
\begin{equation}\label{eq:corrfunc}
 F(\Delta B)=\langle \sigma^\mathrm{el}(B+\Delta B)\sigma^\mathrm{el}(B)\rangle-\langle \sigma^\mathrm{el}(B)\rangle^2,
\end{equation}
where the brackets denote the average over $B$.
In the Ericson regime, this function is predicted to be Lorentzian \cite{PhysRevLett.5.430},
\begin{equation}\label{eq:Ericson}
 F(\Delta B)\propto\frac{1}{1+(\Delta B/\Gamma)^2}.
\end{equation}
Since in this section we are only concerned with magnetic field dependent cross sections, $\Gamma$ has units of magnetic field in our case; it easily converts to energy via $E=-\mu_\mathrm{mag}B$, though. Similarly, we also consider the DOS to be expressed in the magnetic field domain, as in Table \ref{tab:doskrbpartialwave}.

In nuclear physics, where the number of asymptotic channels $N_a\gg1$, the correlations predicted by Eriscon have been nicely demonstrated \cite{RevModPhys.82.2845}. Ultracold atom-molecule collisions as in the present work, on the other hand, are an ideal candidate to investigate the onset of Ericson fluctuations for only a few asymptotic channels. They possess a large enough DOS for the statistical arguments to be valid and -- most importantly -- the number of \emph{relevant} asymptotic channels can be very precisely set by choosing the initial hyperfine states of the colliding particles. Unlike the treatment in nuclear physics where all asymptotic channels are considered open, i.e., $N_o=N_a$, in ultracold collisions only $N_o$ of them remain open at infinite separation of the particles and provide a finite outbound flux as $R\rightarrow\infty$. Therefore, in the context of ultracold collisions, we are concerned with $N_o$ rather than $N_a$. Moreover, by setting the short-range coupling parameter $R_a^{(0)}=1$, we might expect the collision complex to decay with the RRKM rate $\Gamma=\Gamma_\mathrm{RRKM}=N_o/2\pi\hbar\rho$, cf.\ \eq{eq:kRRKM}, where $\rho$ is density of states per magnetic field interval.

\begin{figure}
\includegraphics[width=8.5cm]{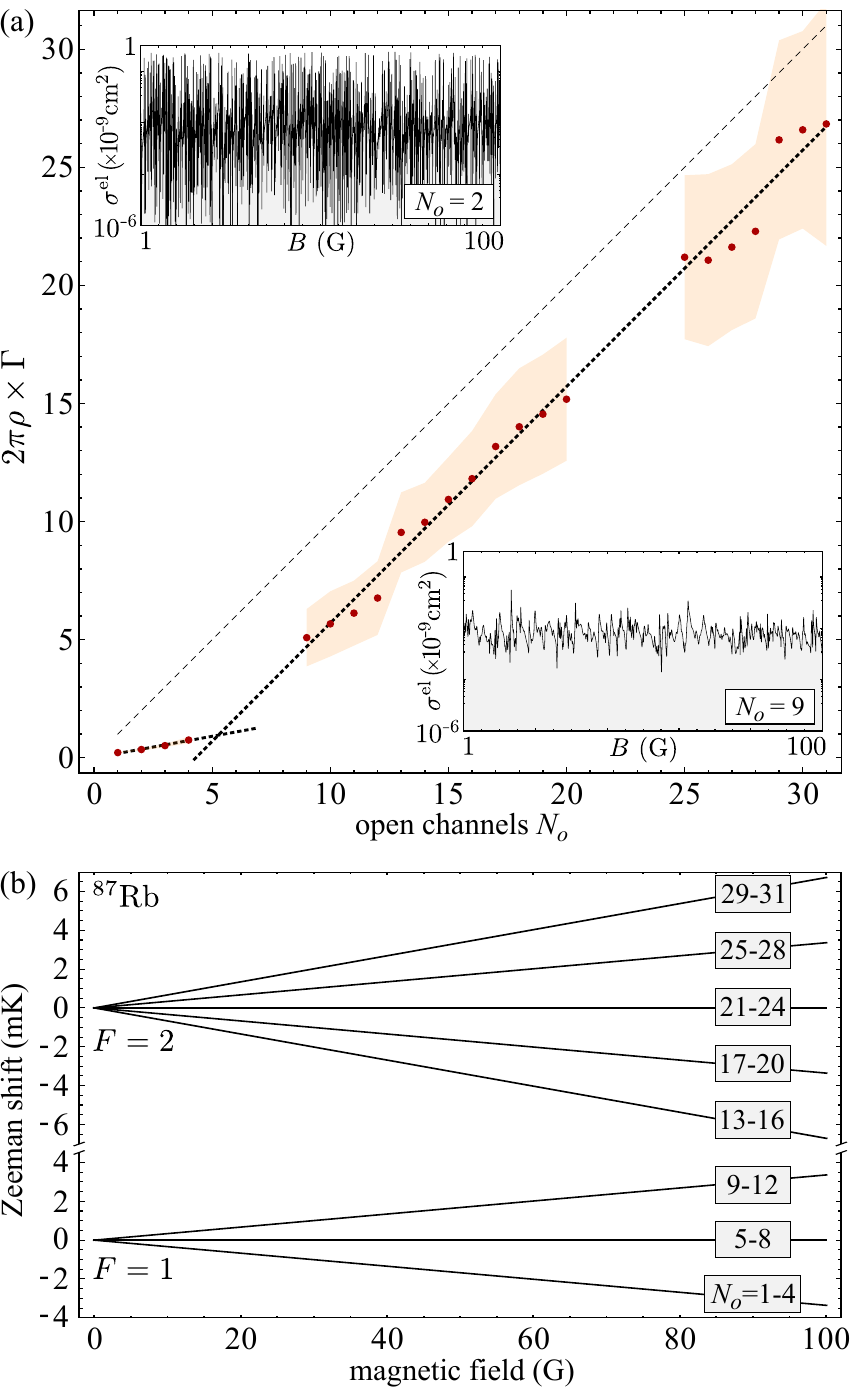}
\caption{(color online) (a) Onset of Ericson fluctuations in ultracold Rb + KRb $s$-wave collisions at a collision energy of 100 nK. Shown is the mean resonance width (dots/solid line) of elastic cross sections as a function of the number of open channels. See text for further details. Insets: Exemplary elastic cross sections for $N_o=2$ and $N_o=9$ open channels, respectively.
(b) Atomic Zeeman shift of the scattering channels. The molecular nuclear spin gives rise to an additional magnetic field dependence which splits each atomic line into a number of sublevels; on the given scale, these sublevels are not visible. The channel indices equal the number of open channels $N_o$.
\label{fig:ericson}}
\end{figure}

In Figure \ref{fig:ericson} we show the mean widths of resonances as a function of the number of open channels. The value of the width is extracted from our scattering data as follows. We consider magnetic field dependent elastic $s$-wave cross sections for Rb + KRb collisions at a fixed collision energy of 100 nK over the range $0<B<100$\,G. The number of open channels $N_o$ is set by varying the incoming channel from the absolute ground state of the system ($f=m_f=1$, $M_\mathrm{K}=-4$, $M_\mathrm{Rb}=3/2$) to the highest possible spin state within the ro-vibrational ground state ($f=m_f=2$, $M_\mathrm{K}=-2$, $M_\mathrm{Rb}=-3/2$). We restrict ourselves to the conserved total magnetic quantum number $M=-3/2$ that contains the absolute ground state, which yields in total 31 possible spin states. After generating a model short-range resonance spectrum $E_\mu$ and width matrix $W_{\mu a}$, we compute for each of the $N_o$ channels the magnetic field dependent elastic cross section. The mean width of the resonances encountered in these cross sections is extracted via their 2-point correlation function \eq{eq:corrfunc} by fitting to the form \eq{eq:Ericson}. This procedure we repeat for 300 randomly sampled short-range resonance spectra and width matrices, for each $N_o$. To accommodate for the different DOS seen by different hyperfine states, we employ $\rho=|m_f|\rho(m_f=1)$. A caveat is that, for states with $m_f=0$, there is a very weak dependence of the incident threshold on magnetic field. In this case, only few resonances are seen in the magnetic field range $0<B<100$\,G, and we cannot extract a width. Thus, in \fig{fig:ericson} no data points are shown for $N_o=5-8$ and $21-24$.

Figure \ref{fig:ericson} presents the results of this simulation, showing the width extracted from \eq{eq:Ericson} as a function of the number of open channels. The dots represent the mean value from 300 trials, while the shaded region indicates the 1-$\sigma$ scatter among the trials. The widths are scaled by $2\pi\rho$ such that the RRKM width \eq{RRKM_rate} is equal to the number of open channels, i.e., it results in a unit slope as a function of $N_o$ (dashed line). The width observed in \fig{fig:ericson} is unquestionably an increasing function of $N_o$, but grows at a different rate for $N_o\le4$ and for $N_o\ge9$.

Focusing on $N_o\le4$ in \fig{fig:ericson}, our data shows a slope significantly smaller than the RRKM value. This can be explained as follows. Assuming that \eq{RRKM_rate} is applicable, resonances begin to overlap when $N_o>2\pi$. Hence, for $N_o\le4$ we are in the regime of isolated resonances where $\Gamma$ is determined by individual widths and not by the collective behavior of many overlapping resonances. Nevertheless, we still find a linear behavior, $\Gamma=\gamma N_o$ whose slope $\gamma$ is determined by the Wigner threshold laws.

The widths of non-overlapping resonances can in principle be extracted from the short-range $K$-matrix $\ksr$. More precisely, \eq{eq:overlap0} yields a mean resonance width in the absence of threshold effects of $\bar{\Gamma}=2/\pi\rho$, which is defined at $R_m$ assuming that particles can freely propagate beyond this point. Threshold effects, which narrow this width, are accounted for within our MQDT treatment. For a single open channel, and neglecting the potential resonant influence of closed asymptotic channels, the MQDT transformation (\ref{eq:mqdttrans}) turns into a simple algebraic equation. Employing further the zero energy limits $\eta\rightarrow 0$ and $G\rightarrow (-1)^{L+1}$ \cite{burkethesis}, the elastic cross section reads
\begin{equation}\label{eq:singlechannelsigma}
 \sigma^\mathrm{el}=\frac{4\pi}{k^2}\frac{A(k,L)^2}{[1+(-1)^{L+1}K_\mathrm{sr}^{-1}]^2+A(k,L)^2}.
\end{equation}
For an isolated resonance in the short-range $K$-matrix, $K_\mathrm{sr}=-(\bar{\Gamma}/2)/(E-E_\mathrm{res})$, \eq{eq:singlechannelsigma} yields a Lorentzian shaped resonance in the elastic cross section with a width of
\begin{equation}\label{eq:gamma}
 \Gamma(N_o=1)=\gamma=A(k,L)\times\bar{\Gamma}=\frac{2A(k,L)}{\pi\rho}.
\end{equation}
The low energy behavior of the MQDT parameter $A(k,L)$ is known analytically \cite{burkethesis,ruzicunpublished},
\begin{align}\label{Aapprox}
A(k,L)^{1/2}&=\frac{R_\mathrm{vdW}^{L+1/2}\,\Gamma(\frac{3}{4}-\frac{L}{2})}{\sqrt{\pi}2^{L-1/2}(2L+1)!!}\,k^{L+1/2},
\end{align}
$R_\mathrm{vdW}$ being the van der Waals length $R_\mathrm{vdW} = (2 m_r C_6 / \hbar^2)^{1/4}$. For the parameters of \fig{fig:ericson} ($s$-wave, $E=100$ nK), we thus find $2\pi\rho\times\gamma=0.18$. This line is shown as a dotted line in \fig{fig:ericson} for $N_o\le4$; it agrees quite well with our numerical result. Hence, even in the limiting case of isolated resonances, $N_o<2\pi$, we find that the RRKM assumption is true: the decay rate and therefore the width of the resonances scales with the inverse of the DOS and is proportional to the number of available outgoing channels.

\begin{figure}
\includegraphics[width=8.5cm]{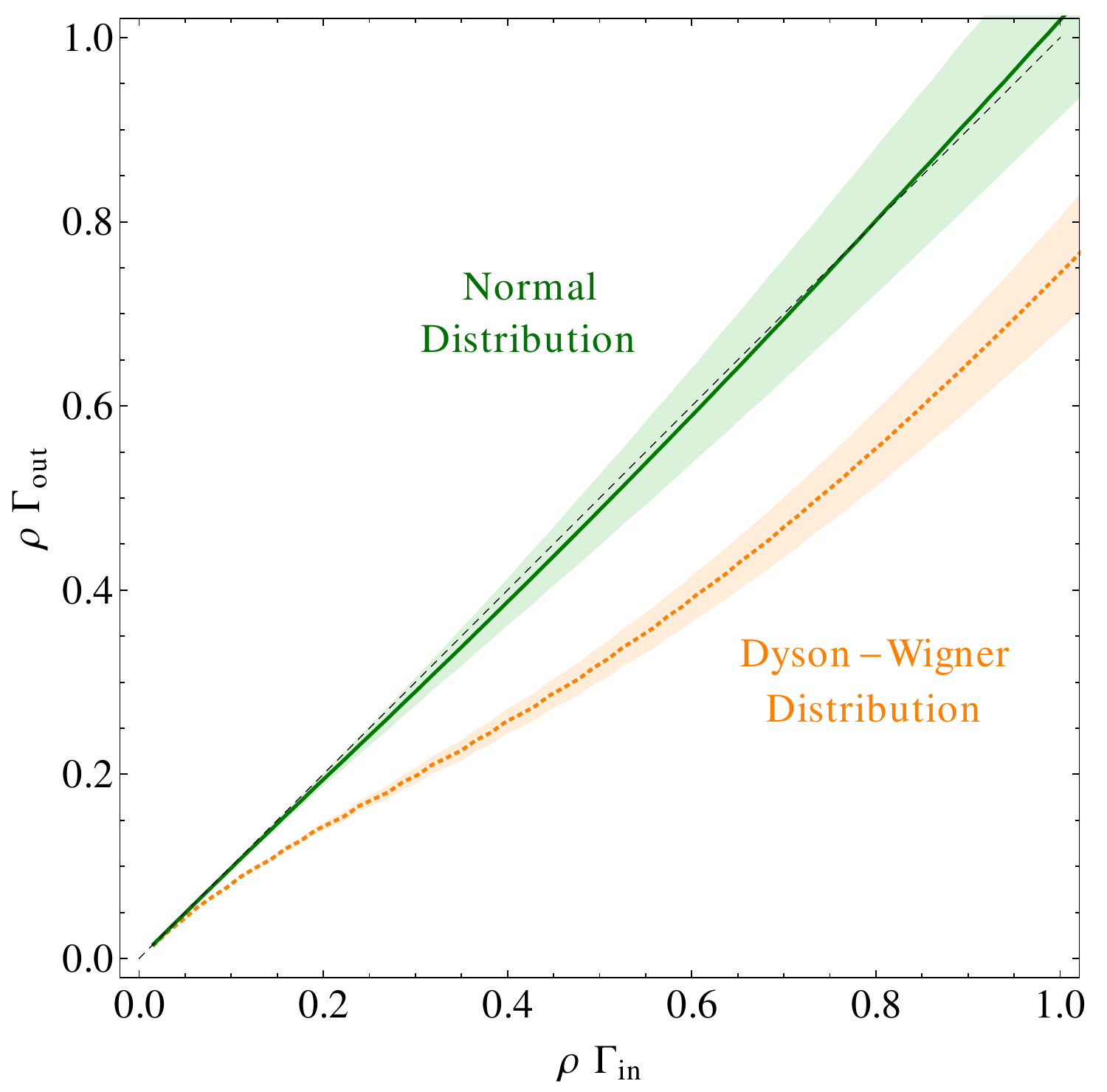}
\caption{(color online) Influence of the Dyson-Wigner distribution on the correlation function \eq{eq:corrfunc}. Shown is the calculated width $\Gamma_\mathrm{out}$ of the correlation function as a function of the input width $\Gamma_\mathrm{in}$ for the Lorentzian model spectrum \eq{eq:modelspectra}. Orange dotted line: Resonances are distributed according to the Dyson-Wigner distribution \eq{eq:goespectrum}. Green solid line: The same resonances are normally distributed. The shaded areas indicate the standard deviation for the given sample of 30 individual runs.
\label{fig:ericsonLorentz}}
\end{figure}

For $N_o\ge9$, the resonances nominally overlap since $\bar\Gamma/d=N_o/2\pi>1$ in the RRKM formula. If this indeed places us in the Ericson regime, then the width $\Gamma$ extracted from \eq{eq:Ericson} should scale according to $2\pi\rho\Gamma=N_o$, i.e., should form a line of unit slope in \fig{fig:ericson}. And indeed this is true, apart from an offset, as seen in the figure (dotted line for $N_o\ge9$). The source of this offset originates in the Dyson-Wigner distribution of level spacings. Qualitatively, the Dison-Wigner distribution discourages levels from being close together. The onset of overlapping resonances is therefore deferred until higher $N_o$.

More quantitatively, it can be understood by employing a simple model. Instead of the actual cross section, we calculate the width $\Gamma_\mathrm{out}$ of the correlation function of a model spectrum that consists of \emph{identical} Lorentzian resonances of width $\Gamma_\mathrm{in}$ at different magnetic field values,
\begin{equation}\label{eq:modelspectra}
 \sigma^\mathrm{mod}(B)=\sum_\mu\frac{1}{\pi}\frac{\Gamma_\mathrm{in}/2}{(B-B_\mu)^2+(\Gamma_\mathrm{in}/2)^2},
\end{equation}
whose locations $B_\mu$ are distributed according to \eq{eq:goespectrum}. We used the same DOS as in \fig{fig:ericson}. The result is shown in \fig{fig:ericsonLorentz} (orange dashed line) as a function of the input width $\Gamma_\mathrm{in}$ of the Lorentzians. For a single resonance, our procedure yields $\Gamma_\mathrm{out}=\Gamma_\mathrm{in}$. For non-overlapping resonances, $\Gamma_\mathrm{in}\ll1/\rho$, this can be seen in \fig{fig:ericsonLorentz} as well. As the resonances start to overlap, however, $\Gamma_\mathrm{out}$ starts to deviate from its linear behavior by showing a smaller width than the input. For a strong overlap, the linear behavior is recovered again, but now with a constant offset from unity. This is the same qualitative the behavior seen in \fig{fig:ericson}. By contrast, if we repeat the model calculation but now for normally distributed resonances instead of Dyson-Wigner (solid green line in \fig{fig:ericsonLorentz}), the result changes quite drastically: the width calculated from the correlation function reproduces the input width. Hence, we attribute the offset of the calculated widths to the particular statistical properties of the Dyson-Wigner distributed resonances.

We remark that the observation of an autocorrelation function with a Lorentzian shape is a necessary but not a sufficient condition for the existence of Ericson fluctuations. As shown, for example, in the case of Helium photoionization, one finds a Lorentzian shape of the autocorrelation function \cite{PhysRevA.84.023402}. Looking at the microscopic processes in more detail, however, this cross section is actually  dominated by only a few, well-defined resonances not in the Ericson regime. The large number of remaining resonances, that would promote the DOS into the Ericson regime, are of too small intensity to contribute to the cross section. In our case, on the other hand, there is no hierarchy in the intensities and all resonances contribute similarly, which makes the autocorrelation function again a good indicator for Ericson fluctuations.


\section{Connection with Experiments}
\label{sec:opportunities}
Our present model of highly resonant scattering makes various assumptions and is by no means meant as a quantitative description. As a qualitative guide, however, it does predict various trends from one molecule to another, and from one internal hyperfine state to another. In this section we summarize what might be gleaned from experimental data as they become available.

\subsection{Density of States}
The first and most obvious measurement would be that of the density of states itself. This is most easily measured, presumably, by the magnetic field variation of a cross section, as in \fig{fig:csbfield}. Even in the highly resonant
case of Rb + KRb that we explored in detail, we still anticipate $s$-wave resonances spaced an average of $\sim 0.1$ Gauss apart, along with perhaps a smattering of higher-partial wave resonances at sufficiently low temperature. In fact, \eq{eq:gamma} tells us that for one open channel $p$-wave resonances at $E=100$\,nK only possess a mean width of $\bar\Gamma=4\times10^{-7}$\,G, which is below typical experimental resolution. For $s$-wave resonances, on the other hand, the required resolution seems experimentally reasonable. For molecules in their ground hyperfine state, the most likely observable would be loss due to three-body processes near each resonance. In this case, one should consider the effect on observables due to the width of the three-body process, a task we have not attempted here. Otherwise, measurements of two-body loss versus field for molecules in their first excited state would supply a reasonable observable.  

Our estimates of the DOS of various collision partners rely heavily on the assumption that the entire phase space allowed by conservation of energy and angular momentum is in fact explored by the collision complex. We have argued above that the DOS is surprisingly weakly dependent on the maximum number of vibrational or rotational states populated (see Table \ref{tab:doskrb}). Nevertheless, our estimates of both the DOS and the lifetime of the complex are almost certainly upper limits. There exists, however, already one experiment that constrains the DOS, namely, collisions of Rb with ground-hyperfine state KRb at sub-$\mu$K temperatures \cite{Ospelkaus2010}. To infer a DOS from this experiment, we assume that there is a universal Rb + KRb collision rate given by the quantum threshold model of Ref.~\cite{PhysRevA.84.062703}, $K^\mathrm{QT}_\mathrm{Rb+KRb}=\pi(2\hbar^2C_6/m_{r,\mathrm{Rb+KRb}}^3)^{1/4}$. Since Rb + KRb collisions are stable against reactive losses, the overall loss rate is given by the above collision rate time the probability that, during the complex lifetime $\tau$, another Rb atom hits the complex and destroys it (the Lindemann mechanism) \cite{levinebook}. This probability is $\tau$ times the Rb + KRb$_2$ rate $K^\mathrm{QT}_\mathrm{Rb+KRb_2}=\pi(2\hbar^2C_6/m_{r,\mathrm{Rb+KRb_2}}^3)^{1/4}$. As a result, the loss rate is a quadratic function of the Rb density $n(\mathrm{Rb})$,
\begin{align}\label{eq:gammaexp}
\Gamma(\mathrm{KRb}) &{}= \tau\, n(\mathrm{Rb})^2 K^\mathrm{QT}_\mathrm{Rb+KRb} K^\mathrm{QT}_\mathrm{Rb+KRb_2}\\
&{}\approx\tau\, n(\mathrm{Rb})^2\times10^{21}\,\mathrm{cm^6/s^2}.
\end{align}
The JILA experiment on KRb emphatically did not measure a quadratic dependence on the loss rate on $n(Rb)$. Nevertheless, we can extract an order of magnitude estimate. From Ref.~\cite{Ospelkaus2010} we infer for a Rb density of $n(Rb)\approx0.6\times 10^{12}\, \mathrm{cm}^{-3}$ a decay rate of $\Gamma(\mathrm{KRb})\approx 20\,\mathrm{s}^{-1}$. This sets a rough upper limit to the complex' lifetime of $\tau\approx10$\,ms. What does this mean for the density of states? Suppose the lifetime $\tau$ is set by the RRKM expression \eq{RRKM_rate}, assuming only one possible exit channel since both the atom and the molecule are in their ground states. The upper limit on the density of states is then $\rho \approx 10^3/\mu\mathrm{K}$. This is a very high density of states, much larger than the $\approx1/\mu\mathrm{K}$ we estimate for this case. Hence, the experiment at least does not contradict our thinking, though it is by no means a measurement of DOS.

\subsection{Resonance widths}
Within the statistical picture we have outlined, there are patterns in the resonance widths as well as in their distribution. A main result, which we believe to be quite general, is the one in \fig{fig:ericson}. Namely, the width $\Gamma$ deduced from the two-body correlation function of a spectrum grows linearly as a function of the number $N_o$ of open hyperfine channels.  The \emph{rate of growth} is, however, different for $N_o < 2 \pi$ and $N_o > 2 \pi$. In the former case, $d\Gamma/dN_o$ is small and is governed by the Wigner threshold law. In this case the widths of the individual resonances may be difficult to extract; observed widths of very narrow resonances are typically set by the temperature, not the intrinsic magnetic-field width of the resonance.

Vice-versa, in the limit $N_o > 2 \pi$, the resonances are not individually resolved anyway. Here the spectrum becomes a varying background of Ericson fluctuations, characterized by the width $\Gamma$ of the two-point correlation function, as in Eqs. (\ref{eq:corrfunc}, \ref{eq:Ericson}). Determining this $\Gamma$ is presumably an easier task, experimentally, than locating and measuring the widths of individual narrow resonances. In such a case, the simplest version of our model predicts that the DOS can be extracted from the slope of $\Gamma$ versus the number of open channels, via $d \Gamma / dN_o = (2 \pi \rho)^{-1}$.

In cases where $\Gamma$ and the DOS $\rho$ are both measured independently, more information can be determined about the microscopic scattering system. Then, as in Fig.\ \ref{fig:ericson}, the dimensionless quantity $2 \pi \rho \Gamma$ can be directly evaluated as a function of $N_o$. Our basic model predicts that this relation will be linear with unit slope, but shifted so as to intercept the $N_o$ axis at some positive value. This shift is, as we have argued, a consequence of the Wigner-Dyson statistics of level spacings. If, on the other hand, the empirical plot of $2 \pi \rho \Gamma$ versus $N_o$ intercepts the origin, the resonances are more likely normally distributed (see Fig.~\ref{fig:ericsonLorentz}). Such data would therefore provide evidence that the underlying classical dynamics of the collision complex is not chaotic.

Furthermore, the experimentally determined $2 \pi \rho \Gamma$ versus $N_o$ may not have unit slope, but may grow more slowly, perhaps even nonlinearly. This would indicate that the widths are narrower than in the ``maximal coupling'' limit we have assumed, in which the coupling parameter $R_a^{(0)}=1$ for all asymptotic channels. Determining realistic values of coupling constants $R_a^{(0)}$ (or equivalently, transmission probabilities $T_a$) would further constrain models of how the ro-vibrational ground state couples to the resonant complex, in ways that remain to be explored.

\subsection{Likelihood of Changing Spins}
Estimates of the spin-changing probability $P$ are arguably the weakest point of our discussion above. Here again, experiments should shed light on the true situation. For example, if molecules are prepared in their second-lowest
hyperfine state, one could simply directly measure the rate at which molecules are produced in the lowest state. When the probability $P$ is of order unity (or higher, in our estimates) the spin-changing rate constant should be on the order of the universal rate constant $K^\mathrm{QT}$, as described above. However, when $P \ll 1$, as we estimate for Na + LiNa or K + LiK, then the spin-changing rate is likely smaller by a factor of $P$. Even an approximate determination of $P$ in this way would be a strong and useful constraint on the time scales that govern the spin-changing dynamics. 

The argument for whether or not the spin changes depends on the DOS. Thus we can also infer useful information from a direct measurement of the DOS itself. Within our order-of-magnitude estimates we expect the density of ro-vibrational states to be nearly universal, i.e., to be something like 0.1 $s$-wave resonances per characteristic van der Waals energy $E_\mathrm{vdW}$. For heavier molecules with $P \sim1$, we expect the density of states to be augmented by the number of spin states, as in \eq{eq:rhohf}. Thus for example if the DOS were measured directly for both K + LiK (where spin does not change) and for Rb + KRb (where spin can change), both expressed in units of resonances per $E_\mathrm{vdW}$, then their ratio would be
\begin{equation}
\frac{ \rho({\rm Rb}+{\rm KRb}) }{ \rho({\rm K}+{\rm LiK}) } \approx
 N_{\rm nuc}({\rm Rb}+{\rm KRb})=31.
\end{equation}
On the other hand, if it turns out that the spin cannot change in Rb+KRb either, this ratio is closer to unity. If instead both collision freely allow spins to change, then the ratio would be closer to $N_{\rm nuc}({\rm Rb}+{\rm KRb})/N_{\rm nuc}({\rm K}+{\rm LiK}) \approx 3.4$. Checking this kind of scaling would provide information for either verifying or refuting our assumptions about whether and how the spin can change.


\section{Conclusions and outlook}
\label{sec:conclusion}
In the present work we have formulated a theory for cold and ultracold atom-molecule collisions that incorporates the ro-vibrational Fano-Feshbach resonances in a statistical manner while treating the long-range physics exactly within MQDT. We provided estimates for the densities of states encountered for all non-reactive collisions involving alkali atoms and heteronuclear alkali dimers. The question if during the collision the hyperfine states of the collision partners are allowed to change is answered by means of a semiclassical approach that estimates the lifetime of the collision complex. As it turns out, we expect all systems except the very lightest ones to allow for such transitions. This has a great influence on the scattering process itself since more resonances are accessible, pushing the cross sections at ultracold temperatures over the limit from showing few to many resonances. Exemplary elastic cross sections as a function of the collision energy are provided for K + LiK (no hyperfine change) and for Rb + KRb (change in hyperfine sublevels allowed). For the latter, also thermalized rates as a function of magnetic field strength for fixed temperature are shown. Since we assume that the short-range physics -- and therefore also the ro-vibrational resonances -- is independent of external fields, the density of states becomes a function of magnetic field that is probed by a rate corresponding to the atomic magnetic moment. This translates the density of states from just a few within the Wigner threshold limit to many per Gauss.

One of the intriguing aspects of the considered ultracold collisions is that the initial states can be very well controlled. This allows one to tune the number of open channels very precisely and opens the opportunity to probe the onset of Ericson fluctuations. The latter are well-known in nuclear physics where one encounters usually a large number of (open) asymptotic channels. Here we showed that the scaling law predicted by Ericson should be nicely observable in ultracold atom-molecule collisions. Moreover, by limiting the number of open channels to a small number, one can switch between being in the Ericson regime and the regime of isolated resonances.

The Ericson fluctuations only depend on the density of states and the number of open channels. Since the latter are fixed by the choice of the initial state, a measurement of the Ericson fluctuations should in principle allow for an experimental determination of the density of states -- an intriguing possibility which is subject of future investigations. Also, in the present work only elastic processes are investigated. The extension to inelastic or chemically reactive ones is straightforward and promises further insights in the physics of highly resonant scattering.

\begin{acknowledgments}
The authors acknowledge financial support from the U.S.\ DOE. M.M.\ acknowledges financial support by a fellowship within the postdoc-programme of the German Academic Exchange Service (DAAD).
\end{acknowledgments}

\bibliography{literature}

\end{document}